\documentclass[acmsmall]{acmart}

\usepackage[T1]{fontenc}

\AtBeginDocument{%
  \providecommand\BibTeX{{%
    \normalfont B\kern-0.5em{\scshape i\kern-0.25em b}\kern-0.8em\TeX}}}

\setcopyright{acmcopyright}
\copyrightyear{2024}
\acmYear{2024}
\acmDOI{10.1145/1122445.1122456}

\acmJournal{TOSEM}
\acmVolume{37}
\acmNumber{4}
\acmArticle{111}
\acmMonth{02}

\usepackage{hyperref}
\usepackage{soul}

\definecolor{chestnut}{rgb}{0.8, 0.36, 0.36}

\usepackage{multirow}

\begin{document}


\title[A Disruptive Research Playbook for Studying Disruptive Innovations]{A Disruptive Research Playbook for Studying Disruptive Innovations}


\author{Margaret-Anne Storey}
\email{mstorey@uvic.ca}
\affiliation{%
  \institution{Department of Computer Science, University of Victoria}
  \streetaddress{P. O. Box 3055, STN CSC}
  \city{Victoria BC}
  \country{Canada}
}

\author{Daniel Russo}
\authornote{Corresponding author.}
\email{daniel.russo@cs.aau.dk}
\orcid{0000-0001-7253-101X}
\affiliation{%
  \institution{Department of Computer Science, Aalborg University}
  \streetaddress{A.C. Meyers Vaenge 15, 2450}
  \city{Copenhagen}
  \country{Denmark}}

\author{Nicole Novielli}
\email{nicole.novielli@uniba.it}
\affiliation{%
  \institution{Department of Computer Science, University of Bari}
  \streetaddress{Via Orabona, 4}
  \city{Bari}
  \country{Italy}
}

\author{Takashi Kobayashi}
\email{tkobaya@c.titech.ac.jp}
\affiliation{%
  \institution{Department of Computer Science, Tokyo Institute of Technology}
  \streetaddress{2-12-1 Ookayama, Meguro-ku, 152-8552}
  \city{Tokyo}
  \country{Japan}
}

\author{Dong Wang}
\email{p.hanel@essex.ac.uk}
\affiliation{%
  \institution{Department of Information Science and Technology, Kyushu University}
  \streetaddress{744 Motooka, Nishi-ku, 819-0395}
  \city{Fukuoka}
  \country{Japan}
}

\renewcommand{\shortauthors}{Storey et al., 2024}

\begin{abstract}
As researchers, we are now witnessing a fundamental change in our technologically-enabled world due to the advent and diffusion of highly disruptive technologies such as generative AI, Augmented Reality (AR) and Virtual Reality (VR). In particular, software engineering has been profoundly affected by the transformative power of disruptive innovations for decades, with a significant impact of technical advancements on social dynamics due to its the socio-technical nature. 
In this paper, we reflect on the importance of formulating and addressing research in software engineering through a socio-technical lens, thus ensuring a holistic understanding of the complex phenomena in this field. We propose a research playbook with the goal of providing a guide to formulate compelling and socially relevant research questions and to identify the appropriate research strategies for empirical investigations, with an eye on the long-term implications of technologies or their use. We showcase how to apply the research playbook. Firstly, we show how it can be used retrospectively to reflect on a prior disruptive technology, Stack Overflow, and its impact on software development. Secondly, we show it can be used to question the impact of two current disruptive technologies: AI and AR/VR. Finally, we introduce a specialized GPT model to support the researcher in framing future investigations. We conclude by discussing the broader implications of adopting the playbook for both researchers and practitioners in software engineering and beyond. 
\end{abstract}




\maketitle

\section{Introduction}

\begin{quote}
``Disruptive technologies aren’t disruptive by themselves, it is how people use them that makes them disruptive''

\begin{flushright}
\textit{---Anonymous}
\end{flushright}
\end{quote}
\vspace{2mm}

As software engineering researchers, our passion lies in understanding and developing theories about software engineering and crafting software engineering solutions. Yet, in our pursuit of scientific and technological excellence, we occasionally lose sight of the broader purpose: the developers, the end users of software, and the societal impact our work is meant to serve. This oversight is not just a personal lapse but a professional one, as our role as scientists demands that we create value for society at large. 
It is easy to fall into the trap of assuming that the research questions that fascinate us will inherently fascinate and serve others. However, this assumption can often lead us astray. 

Recognizing this crucial gap, we have crafted a provocative playbook, offering strategic guidance to those among us who struggle to identify compelling and socially relevant research questions in software engineering. The playbook, we hope, will provoke researchers to pause and to take a step back to reflect on what other questions they could be answering through their research, and what the impact of those answers may be on society at large. We propose a playbook, rather than a framework, as it is a more actionable approach to broadening the research questions we ask. We also introduce ``MyResearchPlaybook'', a specialized GPT model, as an
adjunct to our research playbook, to help in the brainstorming activity the playbook
invokes.  

We justify the need for such a playbook, by tracing the arc of human history. This arc reveals a continuum of technological innovation, from the earliest tools enhancing physical capabilities, to the sophisticated digital systems of today that amplify our intellectual prowess. This relentless pursuit of advancement has been a hallmark of human civilization~\cite{mokyr1992lever}. Each technological breakthrough has brought us forward as a society--from primitive hunting tools to the development of language, writing systems, and beyond--and has fundamentally altered how we work, communicate, and create. These leaps not only showcase our ingenuity but also challenge us to reflect on the role and impact of our creations within the larger societal context.
%
Marshall McLuhan, a visionary scholar active in the 1960s, astutely observed that disruptive technologies follow intrinsic laws: they amplify certain human abilities, render previous technologies obsolete, and when pushed to their extremes, create a need for new innovations to address emergent challenges~\cite{mcluhan77}. McLuhan’s insights compel us to reflect not only on the immediate benefits of new technologies, but also to anticipate their long-term and potentially unforeseen impacts. Consider the automobile: while it has revolutionized transportation, its excessive use has led to traffic congestion and significant environmental degradation, necessitating further innovation and intervention.

The realm of software engineering has not been immune to the transformative power of disruptive innovations since it is fundamentally socio-technical, intertwining technical advancements and social dynamics. Despite rapid progress through innovations like high-level programming languages and social coding tools, the integration of tools such as static analysis and automated debugging remains a challenge. Parnin and Orso's study highlighted that the efficacy of these tools is contingent not only on their technical capabilities, but also (and maybe more importantly) on their adoption and use by developers in real-world scenarios~\cite{parnin2011automated}. This underscores the necessity for a holistic approach in software engineering, ensuring that technical innovations are seamlessly aligned with and supportive of the human elements of software development.

Generative AI technology, along with Augmented Reality (AR) and Virtual Reality (VR), exemplifies the current wave of disruptive technologies showing the potential to revolutionize the field of software engineering. These innovations are not merely incremental improvements; they are catalysts for fundamental change, challenging existing paradigms, and creating new opportunities for advancement and understanding.
The integration of AI is transforming various aspects of software engineering, automating routine tasks, and injecting creativity into software design and development, while simultaneously lowering the barrier to software authorship. Concurrently, AR and VR are emerging as potential disruptors, with technologies such as the Metaverse~\cite{yilmaz2023examining} anticipated to unlock opportunities that enhance e.g., our understanding of complex dependencies. These technologies also play a crucial role in improving collaborative efforts during development and bolstering educational support, as highlighted by Krause~\cite{krause2022collaborative} and Fernandes~\cite{fernandes2022systematic}.

While AI, AR, and VR are the disruptors of today, shaping the current and future landscape of software engineering, it is important to note that they are examples in a long line of innovations. To navigate the ever-evolving landscape of software engineering in our research, we can not overlook the long-term implications of technologies or their use. Thus, the research playbook we propose was intentionally designed to be provocative, versatile, and technology-agnostic. This ensures that it retains its relevance and applicability, providing a reliable guide through the ongoing waves of innovation and disruption in software engineering.

Given the sweeping changes brought about by disruptive technologies, it is imperative for research within this domain to adopt a socio-technical lens, as the lesson of Parnin and Orso shows us, ensuring a holistic understanding that encompasses both the technological innovations and the social contexts in which they operate. This approach goes beyond merely selecting the right research methodologies or adhering to specific methods, as guided by established resources in software engineering~\cite{wohlin2012experimentation,ralph2020acm}. While these guides are invaluable for ensuring methodological rigor, they fall short in assisting researchers to formulate critical questions that delve into the broader and long-term impacts on both technical and socio-technical phenomena in the field.

Here, we underscore the importance of integrating a socio-technical perspective in all studies within this domain. This integration is crucial not only for harnessing the full potential of disruptive technologies, but also for mitigating potential negative impacts, and fostering a sustainable and inclusive technological future. The emphasis here is on encouraging researchers to adopt a holistic perspective that encompasses not just research methods, but also a critical examination of the underlying questions asked and the wider phenomena impacted by disruptive technologies in software engineering. This comprehensive approach ensures a balanced and thorough understanding of the subject matter and its socio-technical implications.

Our paper begins by defining what constitutes a ``disruptive technology'', followed by an overview of key disruptive technologies that have significantly impacted software engineering (Section~\ref{sec:background}). Subsequently, our research playbook is presented in detail (Section~\ref{sec:playbook}), serving as a tool to evaluate innovative technologies in software engineering.
As we present the playbook, we illustrate how the playbook can be used to \emph{retrospectively reflect} on a prior disruptive technology, Stack Overflow, and its impact on software engineering. This retrospective application of the playbook revealed to us how relatively little research emphasis there has been on studying social and human aspects.
To demonstrate practical forward looking applications of the research playbook, we explore two specific disruptive technologies in software engineering that are still emerging: AI and AR/VR. Specific example scenarios are provided to illustrate how the playbook can be employed within these domains (Section~\ref{sec:examples}). In Section ~\ref{sec:examples} we also introduce ``MyResearchPlaybook'', a specialized GPT model to assist in the brainstorming activity the playbook
invokes for the second scenario. We conclude the paper by discussing the broader implications of adopting the playbook, highlighting its potential benefits for both researchers and practitioners (Section~\ref{sec:discussion}).

\section{Disruptive Software Engineering Technologies Disrupt Research}
\label{sec:background}

Frequent instances of technological leaps and creative breakthroughs have a history of shattering existing stable environments, elevating performance across various dimensions, and profoundly reshaping entire societies. These profound shifts are termed ``disruptive technologies'' by Christensen et al.~\cite{christensen:95} This concept has been expanded to encompass innovations not only in technology, but also in products, processes, and business models, known as ``disruptive innovation''~\cite{christensen:00}.

Christensen's original definition of disruptive innovation, which focuses on the competitive dynamics in economic activity, contains two meanings. The first is the disruption of economic market structures, and the second refers to the disruption experienced by incumbent firms. 
Different from sustainable innovations, which operate changes in the market that are predictable by introducing incremental changes in existing business methods, disruptive innovations introduce significant changes in the habits of customers, thus leading to changes in processes and business models that are not predictable in their emergence and evolution. 
Usually, such disruptive innovations emerge from a group of customers whose needs appear not fully addressed by mainstream solutions. The quality of the innovative product or service then improves over time, thus reaching the mainstream market and introducing major changes in the daily habits of people.
Christensen further observed that normally successful and stable businesses fail not at the introduction of sustainable innovations (they are well positioned to stay on top of these changes technology and market wise), but fail in the face of disruptive innovations which often lead to cheaper new products and shifts in new emerging markets~\cite{christensen:95}.  Established companies may be slow to recognize shifts in market needs, may face inertia to change and fail to recognize the potential of the new innovations, and consequently stick with formerly effective processes and tools that are no longer successful in the face of disruptive change. 

The Internet can be seen as an example of a disruptive technology. Originally designed to address the needs of research and military institutions for robust networking technology for collaboration purposes~\cite{ARPANET}, the Internet evolved over time and eventually had a profound impact on individual and collective aspects of our society. It dramatically reshaped the way we communicate with each other, while enhancing our ability to access and share information, and enabling the creation of new business models.  Companies that embraced the internet (e.g., the case of eCommerce), were more successful than those that did not.


This conceptualization of disruptive innovation, originally conceived to model market processes, also holds true for the software engineering domain. 
In software engineering, instances of such transformative breakthroughs are frequent. 
For instance, the introduction of Integrated Development Environments (IDEs) revolutionized the interaction between developer and development support tools. They reduced the friction of using different tools, and they standardized processes and tools across teams. IDE plug-ins replaced the need for standalone tools like Computer-Aided Software Engineering (CASE) tools, which enabled simplified environment setup and tool integration. Social knowledge-sharing platforms like Stack Overflow disrupted the barrier of collaboration and drastically changed the use of traditional documentation, such as FAQs and manuals, and promoted collaborative learning and problem solving.

In the contemporary context, the advances achieved by large-scale language modeling (LLM) and virtual reality/augmented reality (AR/VR) through deep learning, and dramatic advances in computing power are undeniably remarkable. To further enhance the disruptive potential of such technologies, we note that they are accessible to a wider audience compared to technologies introduced in the past, whose benefits were thanks to interaction paradigms such as the use of natural language for LLM-based tools (e.g., Copilot).   

We find ourselves in a dynamic phase of disruptive innovation that has the potential to reshape how we interact with technology and experience reality. Just as formerly successful businesses may fail in the face of disruptive rather than sustainable innovations, researchers also need to consider if they too may suffer from inertia and too much faith in existing approaches for doing research when faced with disruptive innovations.  Disruptive innovations change not only the market, but also the value propositions that need to be understood. 
In response to this observation, our paper offers a comprehensive and action inspiring research playbook that encourages researchers to reflect on the socio-technical phenomena affected by disruptions, such as AI and AR/VR adoption, and the essence of the research questions they should be asking (based on the ideas of how these phenomena may be impacted). We further leverage the disruption of generative AI on research itself by designing a specialized GPT model, called ``MyResearchPlaybook''.  That is, not only do we anticipate that generative AI is disrupting software engineering, but it may also directly disrupt how we do research and disrupt the very questions that we pose in our work.  But first, we present the design of the playbook next. 

\section{The Design of A Disruptive Research Playbook} 
\label{sec:playbook}

We designed and developed a research playbook to guide and assist in the evaluation and prediction of how disruptive innovations may impact software engineering. Frameworks to guide software engineering research have been proposed before~\cite{easterbrook_selecting_2008, wohlin2012experimentation, stol-theories13}, but these focused on selecting and using research methods with little attention to the importance of selecting a pertinent research question. 
Our playbook, in contrast, emphasizes a step by step process, starting with the selection of \textbf{research problems} to be considered, followed by which socio-technical \textbf{phenomena} may be affected by the introduction of a new technology, and which \textbf{ideas} should be considered.  That is, it emphasizes the construction and selection of \textbf{research questions}. Lastly, the playbook guides a reflection on the selection of \textbf{research strategies} to use while studying the potential disruptive impacts of a new technology, with a focus on human and social aspects.

The playbook is inspired by two research frameworks, discussed below. Throughout, we use a former disruptive technology in software engineering, Stack Overflow, which disrupted how developers share, curate, and learn knowledge, to illustrate how the playbook can be used not just to question the impact of new technologies, but also to reflect on the impact and research that has been done on prior disruptive technologies.  

\begin{figure}[tb]
\centerline{\includegraphics[width=0.35\linewidth]{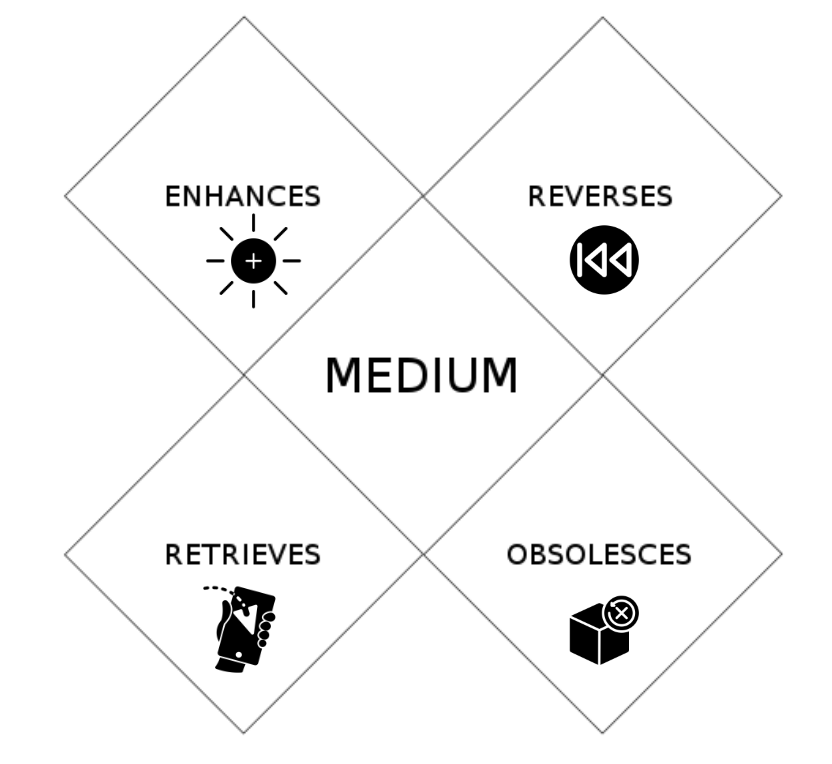}}
\caption{The McLuhan's Tetrad showcasing the four laws that every new medium follows.}
\label{fig:tetrad}
\end{figure}

\subsection{Framing the Research Landscape with McLuhan's Tetradic Dimensions} 

When conducting research, the importance of articulating and crafting one's \textbf{guiding research goals and questions} can never be overstated~\cite{bernstein14, easterbrook_selecting_2008}. At the heart of any empirical research study or evaluation is the \textbf{phenomenon} or phenomena to be studied and the \textbf{ideas} we have about them, while the \textbf{research questions} become the navigational compass points that are considered in the research study.   

Marshall McLuhan recognized some decades ago the importance of choosing probing questions to ask about new media~\cite{mcluhan77} and their impact on humans and society. By media, McLuhan refers to any technology that augments or enhances a human or their capabilities. For example, he considered a car, a poem, a pen, and clothing all as forms of media.\footnote{Although McLuhan used the term media more than technology, we use the more familiar term of technology throughout our paper.}  
McLuhan recognized from his studies of media that any technology, once it is put to use, exhibits the same four laws~\cite{mcluhan2017medium}. These state that a new disruptive innovation will \emph{enhance} or accelerate some aspect of a human's capabilities, it will \emph{retrieve} some characteristics or affordances from older innovations, it will make \emph{obsolete} some existing innovation, and lastly when overused or used to an extreme over time it will \emph{reverse} the very capabilities it strove to provide and eventually flip into a newer technology.  Predicting what will come next is not trivial, and McLuhan also noted that understanding what a technology retrieves requires a deep historical understanding of the technologies that came before it~\cite{mcluhan2017medium}.

McLuhan illustrated these four laws using a tetrad (see Figure~\ref{fig:tetrad}), a visualization to help a researcher understand or even predict the impact a new medium may have. 
This tetrad is often referred to by media scholars as \textbf{McLuhan's tetrad}, and it has been commonly used in recent years to evaluate technological innovations~\cite{mcluhan_operationalizing_2007}.

The tetrad poses four questions to evaluate a new technology: 
\begin{enumerate}
    \item What does the technology \textbf{enhance} or amplify?
    \item What does the technology make \textbf{obsolete}?
    \item What does the technology \textbf{retrieve} that had been obsolesced earlier?
    \item What does the technology \textbf{reverse} or flip into when pushed to extremes or overused?
\end{enumerate}

Table~\ref{tab:mclhuanSE} shows representative examples of past disruptive innovations in software engineering and how McLuhan's tetrad can offer insights on what they enhanced, what they made obsolete, what they retrieved from the past, and what they flipped into over time.

\begin{table*}[]
\centering
\caption{
Disruptive Technologies in Software Development: What they enhanced, what they made obsolete, what they retrieved from prior technologies, and what they flipped into over time. }
\label{tab:mclhuanSE}
\resizebox{\textwidth}{!}{
\begin{tabular}{l|ccccc}\toprule 
\begin{tabular}[c]{@{}l@{}}\textbf{
Disruptive Technology}\\ Date Emerged\end{tabular} & \begin{tabular}[c]{@{}c@{}}\textbf{Machine languages}\\ 1960's\end{tabular}                                  & \begin{tabular}[c]{@{}c@{}}\textbf{High-level}\\ \textbf{Prog Langs, OO, Unix}\\ 1970's\end{tabular}                        & \begin{tabular}[c]{@{}c@{}}\textbf{Internet (Repositories, Libraries,} \\ \textbf{Email, ICQ+Bots,}\\ \textbf{Open Source}) \\ 1980's\end{tabular} & \begin{tabular}[c]{@{}c@{}}\textbf{IDEs,}\\ \textbf{Automated Testing}\\ 1990's\end{tabular} 
 & \begin{tabular}[c]{@{}c@{}}\textbf{Social media,} \\ \textbf{Stack Overflow,} \\ \textbf{GitHub, CI, DevOps}\\ 2000's\end{tabular}   
\\\midrule
\textbf{Enhancement}                                                           & \begin{tabular}[c]{@{}c@{}}Programming,\\ Problem solving\end{tabular}                    & \begin{tabular}[c]{@{}c@{}}Abstractions, \\ Design composition \\ Pseudocode \end{tabular}& \begin{tabular}[c]{@{}c@{}}Distributed dev (merging, reuse), \\ Communication\end{tabular}         & \begin{tabular}[c]{@{}c@{}}Agile,\\ quality, reliability,\\ repeatability\end{tabular}                      & \begin{tabular}[c]{@{}c@{}}Community building,\\ Community knowledge sharing,\\ Faster deployments\end{tabular} \\ \midrule
\textbf{Obsolescence}                                                          & \textit{\begin{tabular}[c]{@{}c@{}}Manual Computations, \\ Natural language, \\ Paper\end{tabular}} & \textit{\begin{tabular}[c]{@{}c@{}}Knowledge of\\ machine\\ programming\end{tabular}}                 & \textit{\begin{tabular}[c]{@{}c@{}}Tarballs and diffs,\\ Lost source code,\\ Duplicate efforts\end{tabular}} & \textit{\begin{tabular}[c]{@{}c@{}}Custom tools,\\ Waterfall processes,\\ Business Analysts,\\ Testers\end{tabular}} & \textit{\begin{tabular}[c]{@{}c@{}}Manuals (technical writers),\\ Systems admins, Emails\end{tabular}}   \\    \midrule
\textbf{Retrieval}                                                             & \textit{Mathematics}                                                                                & \textit{Natural language design}                                                                      & \textit{\begin{tabular}[c]{@{}c@{}}Intellectual Property,\\ Communities of Practice\end{tabular}}            & \textit{\begin{tabular}[c]{@{}c@{}}Standardized \\ environments/process\end{tabular}}                                & \textit{\begin{tabular}[c]{@{}c@{}}Social developers, \\ Collaborative problem solving, \\ knowledge sharing\end{tabular}}                            \\ \midrule
\textbf{Reverse/Flip}                                                                  & \textit{\begin{tabular}[c]{@{}c@{}}High level\\ programming\\ languages, OS control\end{tabular}}   & \textit{\begin{tabular}[c]{@{}c@{}}Collaborative dev\\ tools, open source\end{tabular}}               & \textit{\begin{tabular}[c]{@{}c@{}}IDEs,\\ Automated testing\end{tabular}}                                   & \textit{\begin{tabular}[c]{@{}c@{}}Social coding,\\ automated processes\end{tabular}}                                & \textit{\begin{tabular}[c]{@{}c@{}}Artificial Development \\ (behaviour/social data driven)\end{tabular}}       \\ \bottomrule
\end{tabular}}
\end{table*}

\begin{figure}[tb]
\centerline{\includegraphics[width=.9\linewidth]{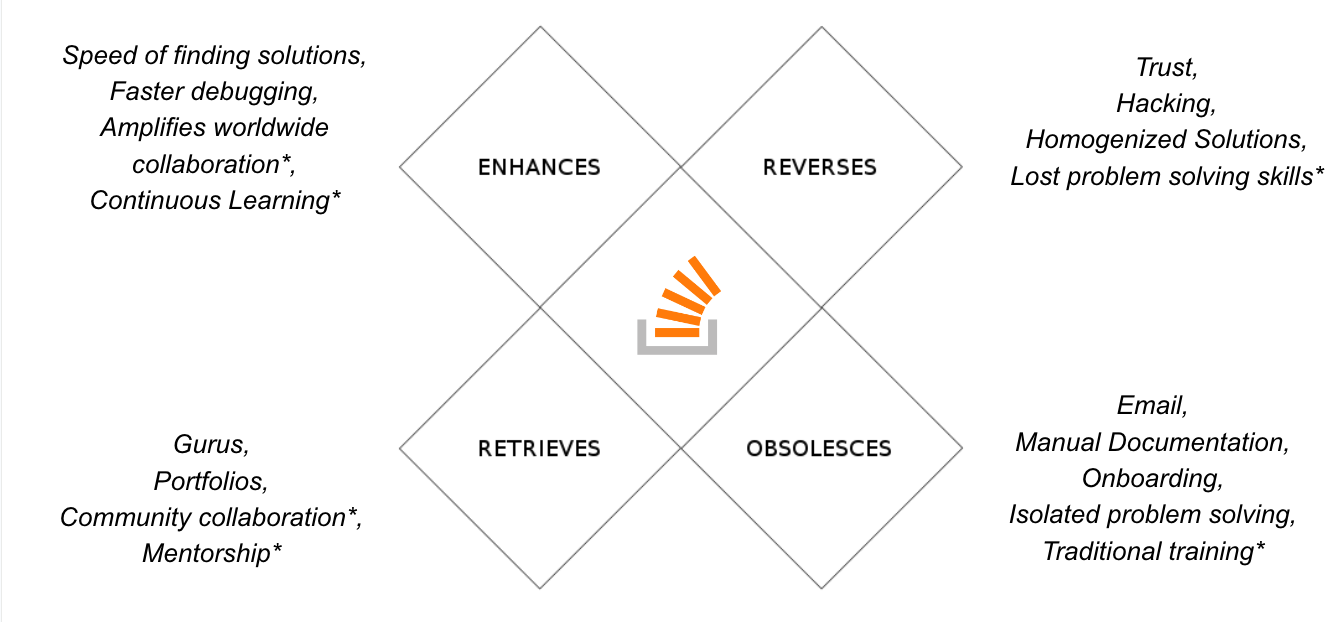}}
\caption{McLuhan's Tetrad applied to Stack Overflow as one example of a disruptive innovation in SE. For each law, we reflect on how Stack Overflow has had an impact on software developers and their practices. Impacts noted with an asterisk were suggested by ChatGPT 4 on Oct 16, 2023.}
\label{fig:tetradstackoverflow}
\end{figure}

McLuhan's tetrad is used as the first step in our playbook for evaluating new technologies in order to reflect on the impact of prior disruptive technologies (such as Stack Overflow). 
Figure~\ref{fig:tetradstackoverflow} shows how McLuhan's tetrad can be used to reflect on the impact that Stack Overflow has had in software engineering. 
Researchers have studied how Stack Overflow has \textbf{enhanced} (sped up) the task of finding answers to technical questions during coding and debugging tasks~\cite{Vasilescu2013Associations,Bacchelli:etAl:IDE,SILVA2021111063,CAMPOS:2016-bugFixing, Lotufo:etAl:BugTracking}, how it has enhanced how developers across the world ``collaborate'' by crowdsourcing this knowledge~\cite{Treude:etAl:AskingQuestions,Mamykina:etAl:designLessons,Barua:etAl:SOTopics,Grant:Buddy:gamification,Asaduzzaman2013Answering,Bosu2013Building}, and how it has been effectively leveraged for retrieving technical information~\cite{Cao:etAl:EfficientSearch} and supporting learning~\cite{Subramanian2013Making,CHATTERJEE2020110454}.
Researchers have also investigated how Stack Overflow has made other technologies (such as email) \textbf{obsolete} or less important~\cite{Squire:moving_to_SO}, and how it has \textbf{retrieved} characteristics from prior technologies, such as the support of mentors~\cite{Ford_etAl:mentoring}.  

What Stack Overflow is \textbf{reversing} into is, however, an important and ongoing topic for research. While code snippets from Stack Overflow answers might be seen as a crucial part of developers' knowledge, their effective reuse is far from trivial~\cite{Yuhao:etAl:2019:understanding:reuse,Treude:Robillard:2017:understsandingSO}. Some researchers have concerns that developers use code snippets from Stack Overflow without
fully understanding how they work, potentially lowering code quality~\cite{MELDRUM2020102516,Digkas:etAl:2019:reusingSO,Ragkhitwetsagul:et:AL:toxicCode:SO,Zhang:etAL:reliableSO:2018}, introducing bugs~\cite{ABDALKAREEM2017148} and security risks~\cite{Yasemin:security:2016,Fischer:security:2017}. There is also a growing concern that the quality of answers is in general decreasing over time~\cite{Zhang:etAL:reliableSO:2018}. 
Because of these problems related to quality, and the challenges of customizing snippets from Stack Overflow, developers have a desire for not just finding solutions, but finding solutions that are personalized to their needs/products.  
Some recent research has investigated how Stack Overflow is ``flipping'' into generative AI solutions, and how generative AI has the potential to reshape software engineering practice~\cite{Ebert:Louridas:2023} and replace community-based question answering~\cite{Burtch:GPTvsSO:2023}.

%
%
McLuhan's tetrad is a provocative tool to guide reflective and predictable research, but more guidance is needed to help decide which phenomena (or actors) and conceptual ideas should be the focus of a study, and how to study these phenomena and ideas.  The next part of the playbook provides guidance on these decisions. 

\subsection{Identifying Research Phenomena Using McGrath's Research Framework}

An essential facet of research design, particularly relevant to software engineering studies, is the coherent structuring of research phenomena, theoretical ideas or constructs, and methodological choices. A pivotal framework that encapsulates these components is Joseph E. McGrath's triadic domain configuration, comprised of the substantive, conceptual, and methodological domains~\cite{mcgrath1995methodology}. Here, we detail each of McGrath's domains and illustrate their application in the realm of innovative research and software engineering, forming the second part of our playbook for studying disruptive innovations through software engineering research. We continue to reflect on prior research about Stack Overflow to illustrate the playbook. 

\subsubsection{The Substantive Domain: Embodying Research Phenomena in SE}
The substantive domain serves as the research empirical backbone, forming the focus of the questions posed. It focuses on the specific phenomena or subjects that prompt academic curiosity and investigation. The \textbf{phenomena} are typically the units of analysis in any research study that help frame the focus of the study. McLuhan's questions about how a technology enhances, makes obsolete, retrieves, and reverses should be asked with these phenomena in mind.  Some example phenomena that have been the focus of research on Stack Overflow include: 

\begin{itemize}

\item Software Artifacts: 
Looking through libraries of software engineering research, we see that there is extensive research on Stack Overflow that studies software artifacts as the unit of analysis, notably questions and the quality of answers, for example see ~\cite{barua2014developers}.

\item Individual Developers: 
Although much of the research on Stack Overflow tended to focus on the quality or content of questions and answers, some researchers considered the direct impact on developers, for example when considering the emotions developers experience using Stack Overflow~\cite{novielli2014towards}. 

\item Software Teams: 
Ponzanelli et al. considered teams as the unit of analysis when they studied how integrating Stack Overflow in the IDE may support how teams better support each other~\cite{ ponzanelli2013seahawk}. 

\item Software Projects: 
Research by Squire et al. studied how companies made the decision to move or not move to Stack Overflow from other channels~\cite{squire2015should}. 

\item Society: 
A societal concern of software engineering is to increase diversity and support inclusion. Research by Ford et al. investigated how the design of Stack Overflow may have introduced barriers and led to less participation by some genders~\cite{ford2016paradise}.


\item Software Processes and Methodologies: 
Stack Overflow is perceived as an invaluable and authoritative knowledge base by software developers, thus fostering the practice of code reuse from discussion threads on the Q\&A site. This phenomenon motivated researchers to investigate Stack Overflow's contribution to the evolution of software processes. 
For example, Tang and Nadi~\cite{Tang:Nadi:2021} investigated how recommendations gleaned from Stack Overflow comment-edit pairs influenced software maintenance practices. 


\item Communities: 
Several studies about Stack Overflow investigated how using it supported crowdsourcing of documentation at a community level~\cite{barzilay2013facilitating,moutidis2021community}.
\end{itemize}

\subsubsection{The Conceptual Domain: Developing Ideas and Theoretical Constructs}
Progressing from the empirical realm of the substantive domain, the researcher enters the \textbf{conceptual domain}. 
This is the domain where new ideas are generated and built upon, and where theoretical constructs are leveraged and developed to enable a deeper understanding of the complexities inherent to the substantive domain. Stol \emph{et al.} described the role of theory in software engineering research and discuss how to leverage theories in software engineering research~\cite{stol-theories13}, while Sjøberg \emph{et al.} described how to build on and extend theories in software engineering~\cite{Sjoberg-theories2008}. More recently, Lorey et al., catalogued how social science theories such as the Technology Acceptance Model, the diffusion of innovation and coordination theories, have been used in software engineering research~\cite{lorey-theories22}. These and other theories provide fodder for ideas that can be explored in software engineering research that considers the disruptive impact of innovative technologies on pertinent phenomena.  

Returning to our illustrative example, research on Stack Overflow has led to new taxonomies to capture the kinds of programming questions asked and answered~\cite{Treude:etAl:AskingQuestions}. Other studies have integrated or built on theories to explain their research insights on effective community-based knowledge creation and sharing. Ford \emph{et al.} provided empirical evidence of the effectiveness of mentoring strategies to support question answering in Stack Exchange~\cite{Ford_etAl:mentoring}; Calefato and colleagues provided empirically-driven guidelines for effective question writing in Stack Overflow. Beyond the question quality, they also investigated the role played by emotions while writing questions and suggest a neutral style increases the likelihood of having a useful answer~\cite{Calefato:etAl:2018:howtoask}.

\subsubsection{The Methodological Domain: Guiding Research Strategies}
The \textbf{methodological domain} attends to the practical dimensions of conducting research. It involves the identification and application of appropriate methods and strategies for data collection, analysis, and interpretation. This domain serves to connect the insights gained from the substantive and conceptual domains with the practicalities of empirical investigation.

\begin{figure}[tb]
\centerline{\includegraphics[width=0.8\linewidth]{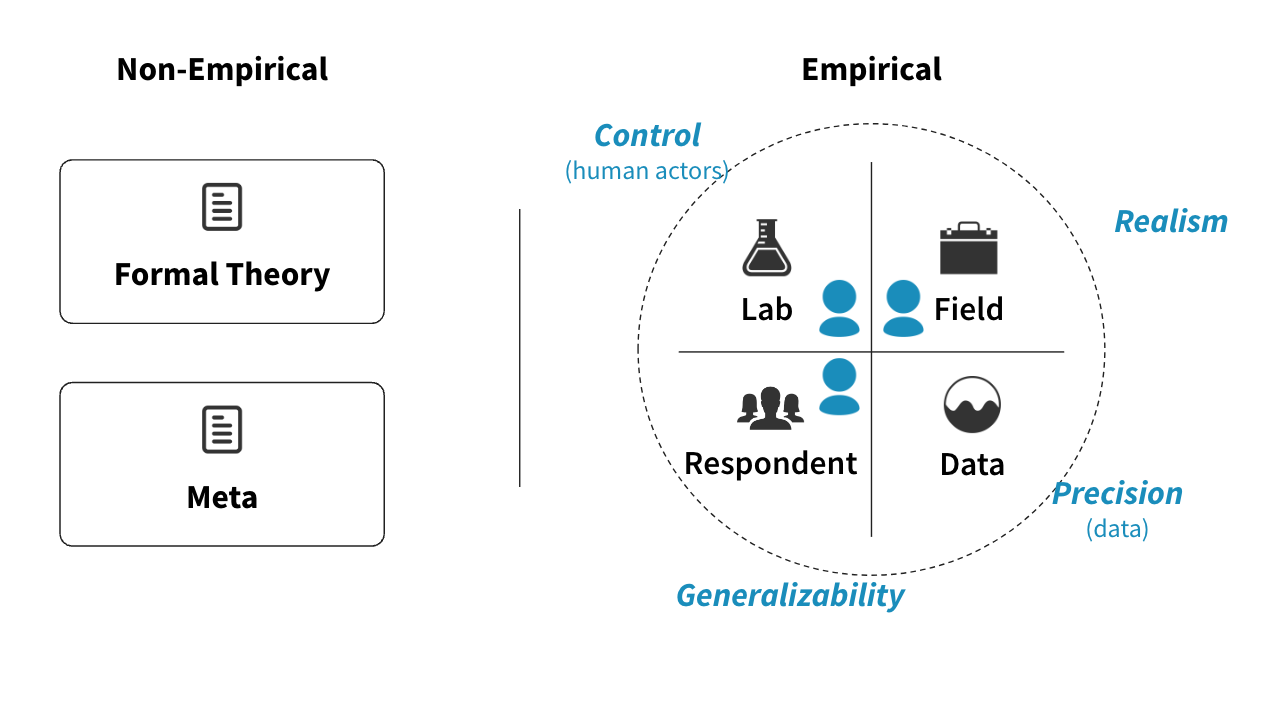}}
\caption{Methodological Domain: Empirical and non-empirical research strategies. Empirical strategies (on the right) are annotated with the quality criteria they have potential to maximize, and which strategies involve direct involvement of human participants (all except data strategies) with the blue person icon. The quality criteria shown on the outside of the empirical circle reveals the inevitable tradeoffs made when a particular strategy is selected.}
\label{fig:www-how}
\end{figure}

There are many resources in software engineering to guide the choice of research methods~\cite{easterbrook_selecting_2008, stol-theories13, storey2020software}. We note that McGrath's circumflex, in its original form, considers human and social aspects as a core consideration in the strategy classification it presents--aspects that are often overlooked in SE research. McGrath's framework, with its focus on human and behavioural aspects, was customized by Storey et al.~\cite{storey2020software} for software engineering research contexts to account for the prevalence of data studies (data mining and simulation studies) in SE.  This extension is referred to as the ``Who, What, How Framework'' and it summarizes research strategies into four broad categories: laboratory studies and experiments (strategies that increase the ability to control the behaviour of human subjects), field studies and experiments (strategies that have the potential to increase realism), respondent strategies (that have the potential to increase generalizability), and data studies and experiments (that also have potential for generalizability and high data precision). The first three strategies in Figure~\ref{fig:www-how} are annotated with a blue person icon to emphasize strategies that directly involve human participants as the unit of observation. 

\subsection{The Research Playbook}
\label{sec:applying}

McGrath's triadic domain framework, and Storey et al.'s extension for SE, underlines the interconnected nature of empirical phenomena, theoretical ideas and constructs, and methodological choices in research design.  By grounding research investigations in McGrath's triadic domain framework and McLuhan's tetrad, researchers can enhance the coherence and robustness of their studies, generating meaningful insights that contribute to the advancement of software engineering. 

\begin{figure*}[t]
\centerline{\includegraphics[width=1\linewidth]{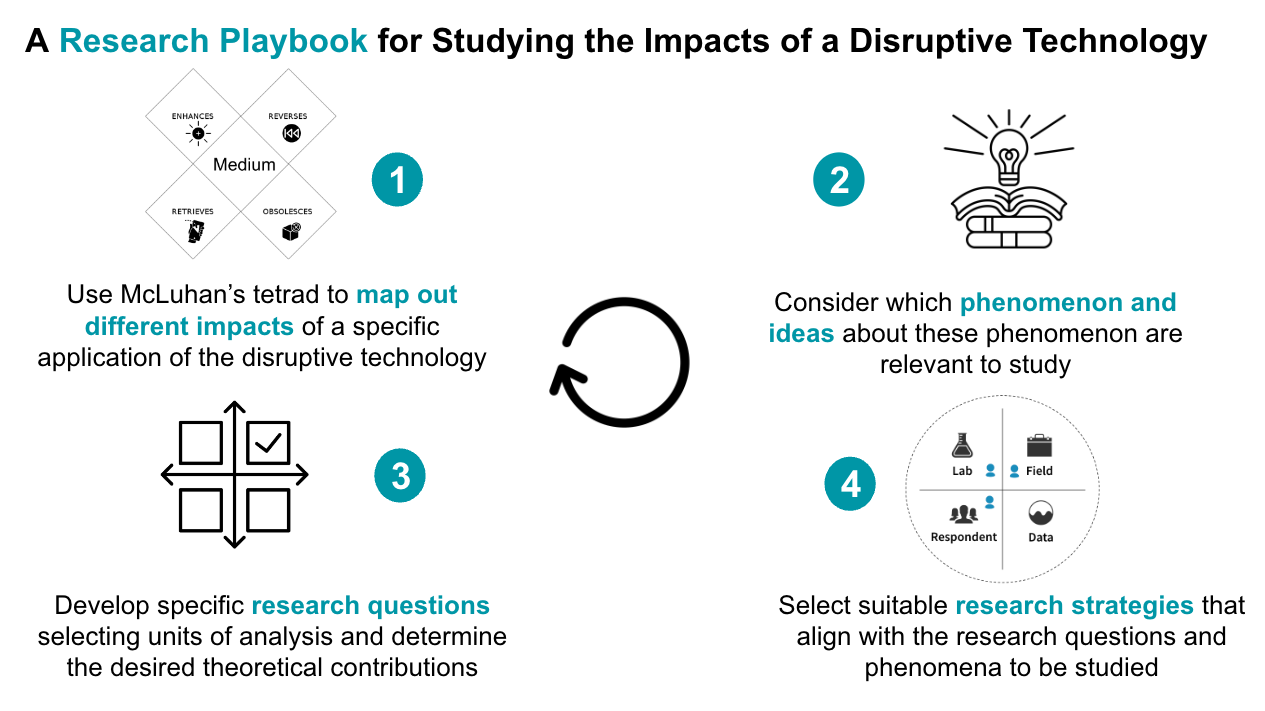}}
\caption{A research playbook for evaluating the impact of disruptive technologies. Step 1 builds on McLuhan's tetrad to brainstorm broad questions. Step 2 identifies phenomena and ideas to study. Step 3 creates a matrix of research questions. Step 4 suggests research strategies that focus on human aspects. The steps may be followed iteratively, refining phenomena, ideas and questions in terms of their potential research impact.}
\label{fig:playbook}
\end{figure*}

In Figure~\ref{fig:playbook} we show the research playbook we propose. We use the term playbook rather than framework because of the step-by-step guidance it provides on the research process to follow, described below. Although we number and present the steps in order, it is important to note that iteration may and often should occur throughout the process of designing a research program or projects. Below, we elaborate further how the playbook can be used to reflect on prior research  As an illustrative example, we used the research on the use of collaborative platforms, such as Stack Overflow, in different areas of software engineering.   

\begin{itemize}
    \item Step 1: Disruptive technologies are only disruptive when they are put to \emph{use} by people for some application.  For this step, McLuhan's tetrad is used to sketch probing questions inspired by the four laws of new technologies. This first step encourages a researcher to step back and reflect on the different possible impacts of a disruptive technology in software engineering. 
    \item Step 2: For this second step, careful attention is paid to which phenomena should be studied in turn guiding the units of analysis for the study, but also which \emph{ideas} we have about the impact of the technology on those phenomena. These ideas may come from experience (as software developers ourselves), or they may emerge from other theories in or outside of software engineering. Choosing phenomena that relate to human and social aspects of software engineering is encouraged. 
    \item Step 3:  In this third and critical step, the researcher selects, crafts, clarifies, and justifies their \emph{research question(s)} considering carefully the different possibilities that the combination of steps 1 and 2 may bring to light. For this step, we suggest creating a table or matrix combining overarching questions posed by McLuhan's tetrad as rows, and the possible phenomena as columns.  Cells in the matrix build on ideas (insights) we may have about the phenomena to be studied. These ideas may come from existing research or from knowledge we have from our own experiences.
    This crucial step of brainstorming and selecting research questions involves further contemplation about the potential impact or actionability of the results from studying these questions, as well as identifying the \emph{type} of and novelty of the knowledge contribution that may arise, asking is it more understanding about the problem domain, or the design or refinement of a technology, or evaluation of the technology~\cite{engstrom2020software}.  
    \item Step 4: For each research question (if more than one), this step involves selecting and justifying research strategies that ideally directly involve human participants. 
\end{itemize}

\begin{figure}[b]
\centerline{\includegraphics[width=.75\linewidth]{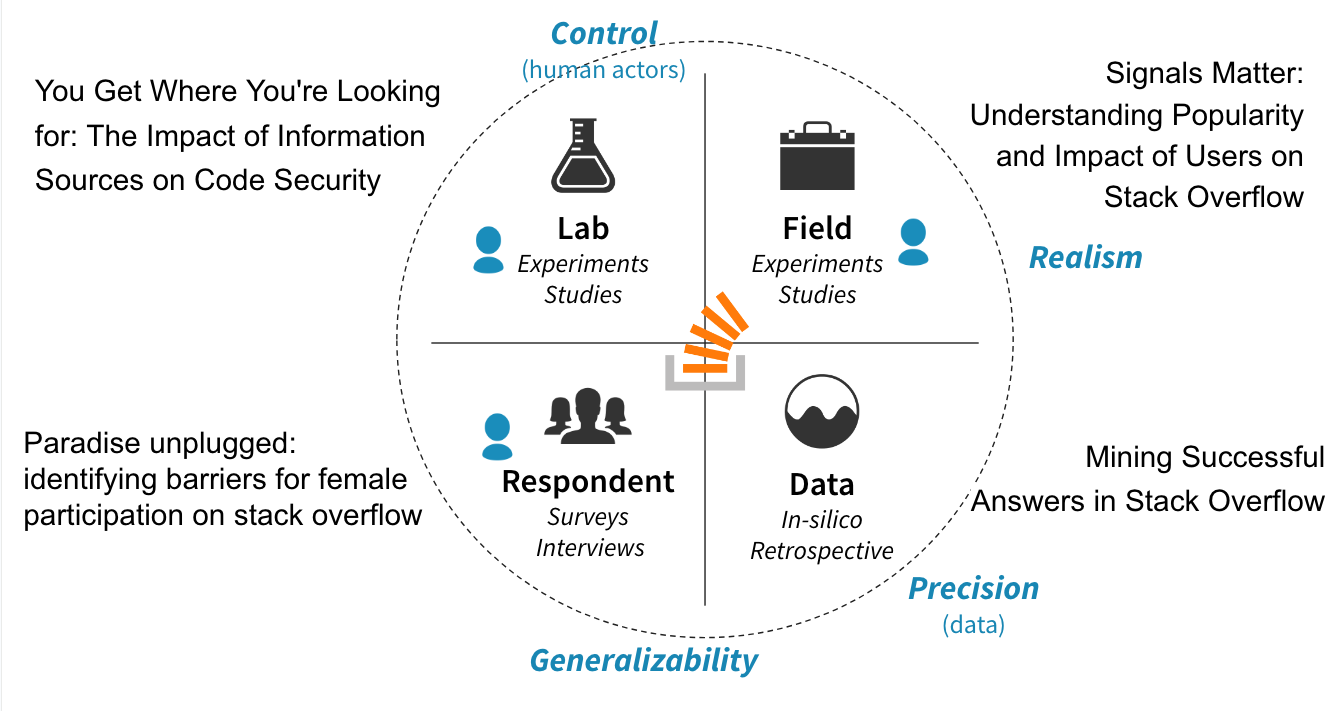}}
\caption{Step 4: This figure shows how we can use this part of the playbook to reflect back on the choice of research strategies to select research studies. Three of these strategies as mentioned directly involve human subjects, but we noted that most studies on Stack Overflow relied on data strategies (many trading precision of data and generalizability for control over human subjects).}
\label{fig:stackoverflow-step4}
\end{figure}

In Table~\ref{tab:stackoverflow-step3} and Figure~\ref{fig:stackoverflow-step4} we show how the playbook is used to reflect on prior research on the impact of Stack Overflow on software development. Table~\ref{tab:stackoverflow-step3} shows a matrix of research questions (each cell is a question, guided by the tetrad) that was already asked about the phenomena of documentation and programmers. Other phenomena could have been selected for consideration here (for example, teams or tools). 
In Figure~\ref{fig:stackoverflow-step4} we show the different research strategies applied by the noted prior research on Stack Overflow. Each choice is a \textbf{trade off}. The choice of a user study means control over human subjects was possible~\cite{Acar:SO:codeSecurity:2016}. A respondent strategy (used by Ford et al.~\cite{ford2016paradise}) increased the generalizability of their results on barriers to using Stack Overflow, while the field study used by Merchant et al.~\cite{Merchant:popularity:2019} on signals in Stack Overflow increased the realism of their study.  The choice of a data mining strategy by Calefato et al.~\cite{Calefato:successfulAnswers:2015}, had high data precision (making it potentially easier to replicate but at the expense of control over human subjects and of realism).  

Using the playbook to reflect on a prior disruptive technology, helped us see some patterns in prior research.  Notably, most research focuses on the enhances dimension, with some research on what it made obsolete. Less research (at least so far) has considered McLuhan's retrieve and reverse dimensions. 
Also we noticed, that much of cited research on Stack Overflow did not consider human and social aspects, but rather emphasized technical aspects in their work. This insight is also corroborated by systematic literature studies on Stack Overflow, such as the study by Meldrum \emph{et al.}~\cite{meldrum2020}.

\begin{table}[t]
\centering
\begin{tabular}{p{2cm}|p{5cm}|p{5cm}|}
\cline{2-3}
                               & \multicolumn{2}{|c|}{\textbf{Phenomena to Study (Stack Overflow)}}                            \\ \hline
\multicolumn{1}{|l|}{\textbf{Triadic Dimensions}}  & \textbf{Documentation}                                           & \textbf{Individual Developers}                                            \\ \hline
\multicolumn{1}{|l|}{\textbf{Enhances}}  & How can Stack Overflow be used to augment API documentation?~\cite{treude2016augmenting}                       & Does StackOverflow enhance the skills of student developers?~\cite{bhasin2021student}  \\ \hline
\multicolumn{1}{|l|}{\textbf{Retrieves}} &  How does the SO voting feature from sites such as Digg impact documentation quality?~\cite{squire2014}    & How do badges (from the realm of boy scouts) steer user behaviour in Stack Overflow?~\cite{anderson2013badges} \\ \hline
\multicolumn{1}{|l|}{\textbf{Obsolesces}} &  Does Stack Overflow replace the need for formal API documentation?~\cite{parnin2012crowd}   &  Can a Stack Overflow Bot replace a formal onboarding process in open source?~\cite{Dominic2020} \\ \hline
\multicolumn{1}{|l|}{\textbf{Reverses}}  & Does Stack Overflow lead to incomplete documentation over time?~\cite{parnin2012crowd} &  Does Stack Overflow, if overused, negatively impact student  learning?~\cite{poial2021} \\ \hline
\end{tabular}
\caption{Step3: This shows various research questions that formerly investigated how Stack Overflow impacted two phenomena: documentation and developers.}
\label{tab:stackoverflow-step3}
\end{table}

In the next two sections, we show how the playbook can guide future research. We delve into two pivotal scenarios that epitomize disruptive innovations in software engineering. These scenarios serve as practical illustrations of how our proposed Playbook can be instrumental in guiding and enriching the investigation of emerging technologies in software engineering practices.

\section{The Playbook in Action}
\label{sec:examples}

To demonstrate the utility and versatility of the Playbook, this paper will focus on two revolutionary technologies that are redefining the landscape of software engineering. The first scenario explores the burgeoning role of large language models, particularly generative AI, which are rapidly evolving and increasingly influencing current development practices. This disruption presents a unique opportunity to observe and analyze a disruptive technology as it emerges and integrates into the fabric of software development.

Moving forward, the second scenario examines the potential of augmented and virtual reality (AR/VR) in software engineering. Despite years of development, AR/VR technologies stand on the brink of a paradigm shift in how software is developed and interacted with. The following section will not only explore the implications of AR/VR but will also demonstrate how tools like ChatGPT can be harnessed for brainstorming and conceptualizing research studies in software engineering.

\subsection{Applying the Playbook to the Disruption of generative AI in Program Synthesis}


In the past several years, there has been a growing interest in harnessing the power of artificial intelligence (AI) and machine learning (ML) technologies to enhance and automate various aspects of the software development life cycle~\cite{barenkamp2020applications}. For instance, code completion, a feature that has been embedded in integrated development environments (IDEs) for years~\cite{murphy2006java}, assists developers by anticipating their intentions and providing relevant recommendations while coding~\cite{wagner2018systematic}.

We are also witnessing a pivotal moment in the progression of AI-supported tools, as systems like Copilot~\cite{copilot2021github}, ChatGPT-4~\cite{openai2023gpt4}, or LLaMA~\cite{touvron2023llama} herald the arrival of a new epoch characterized by disruption and innovation within the field. With its sophisticated capabilities, Copilot can decipher code context and semantics from minimal input, subsequently producing suggestions for not only the next few lines but also entire functions in certain instances~\cite{chen2021evaluating}. The extraordinary user experience furnished by these state-of-the-art large language models (LLMs) is exemplified by their extensive capabilities, superior response quality, and the significant potential to enhance user productivity~\cite{ross2023programmer}. Even users initially skeptical have been won over by the transformative influence of these contemporary LLMs on the software development process~\cite{ross2023programmer}. 



In the following, we show how the playbook can help guide and inform research aimed at systematically evaluating the ramifications of generative AI within the complex socio-technical ecosystem of software engineering. 

\paragraph*{Step 1: McLuhan’s tetrad is used to brainstorm the impacts of applying LLMs on program synthesis.}

The important point of using McLuhan's tetrad is to broaden the questions that may be explored in the research.  In the following, we show one possible iteration of this step in the playbook. Of course, depending on the context of the researcher and research done to date, what comes to mind in this step may vary greatly.  

\begin{itemize}
    \item \ul{In what ways will LLMs \textbf{enhance} the process of writing code?} Faster coding, tests and improved documentation are likely outcomes of using LLMs. In addition to better productivity \cite{Imai-2022-copilot:ICSE22}, quality may also be enhanced and could lead to improved time estimates and faster onboarding. 
    \item \ul{What will LLMs make \textbf{obsolete} when used for program synthesis?} LLMs may make tools like Stack Overflow obsolete, reduce the need for search, reduce the need for manually created documentation, and even reduce demands for education about how to code.  
     \item \ul{What might LLMs \textbf{retrieve} that had been obsolesced before?} LLMs bring back the ability to effectively explain a solution in natural language (see \url{https://github.com/features/copilot}), a more common skill from the earlier days of programming. 
    \item \ul{What will LLMs \textbf{reverse} if used to an extreme for program synthesis?} Although LLMs are thought to improve the abilities of developers, they may over time reduce their abilities should developers trust their output and not spend time understanding the code generated.  They may have less understanding over time of why and how the code works. Furthermore, a blind trust could lead to lower quality code, vulnerabilities, and more bugs that are harder to identify and fix, also due to LLM \textit{hallucinations}, which is a well-known limitation of generative AI. 
\end{itemize}

\begin{figure}[tb]
\centerline{\includegraphics[width=0.9\linewidth]{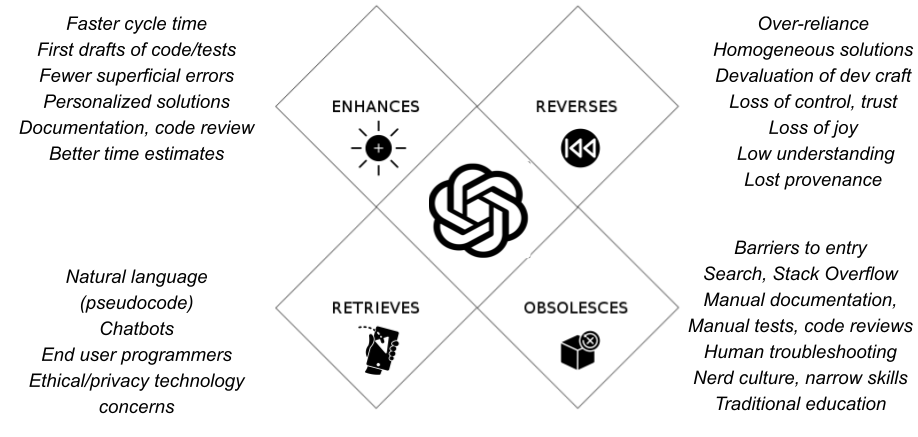}}
\caption{Step 1: McLuhan’s tetrad applied to generative AI in program synthesis to consider many possible impacts.}
\label{fig:tetrad:AI}
\end{figure}

Figure~\ref{fig:tetrad:AI} shows how McLuhan's tetrad is used to assist in brainstorming the impacts generative AI may have on program synthesis capabilities and to consider broader side effects, such as loss of developer skill writing code. Using McLuhan's tetrad as a lens helps to provoke impacts that did not immediately come to mind when we think about how LLMs will only enhance program synthesis speeding up development time.
We note that recently published research focuses mostly on how generative AI enhances or makes obsolete former tooling in SE, with less emphasis on what it will retrieve (such as use of natural language) or reverse into (such as over reliance and poor quality). 

\paragraph*{Step 2: Deciding which phenomena to study and identifying which ideas we may have that we can build on.}

The possible research impacts that emerged from the first step helps to suggest \textbf{phenomena} and \textbf{ideas} that can be studied, with attention paid in particular to human and social aspects, as guided by the intention behind the playbook. 
The phenomena and ideas that become the focus of study are naturally shaped by the prior research and experiences and interests of the researcher. There is already substantial research on the impacts of LLMs on program synthesis and much of this research has focused on understanding the impact of generation on program code quality and accuracy~\cite{hou2023large}. There is also some research that has investigated the impact on developers, e.g., the good day project at GitHub studied how the use of copilot impacted the productivity and satisfaction of developers using Copilot~\cite{Peng:copilot:controlledStudy}.
Our playbook aims to nudge consideration of human and social aspects, thus we consider the impact of this disruptive technology on both individual developers and teams during this application of the playbook. For this step, we also need to consider what other research has been done, but also look to what signals are coming from industry about how the technology is having an impact. 

\paragraph*{Step 3: Creating a matrix of possible research questions concerning phenomena and ideas that can be studied.}

Researchers often commit prematurely to research questions -- this step of the playbook slows down that process and encourages a more careful consideration of the impacts, inspired by the tetrad, crossed with possible phenomena and ideas about those phenomena to brainstorm and inspire new research questions.  As mentioned above, here the playbook encourages us to select and justify specific research questions for further study.  Each cell contains a potential research question to study.
The rows of this matrix correspond to the McLuhan laws (the tedradic dimensions, as mentioned above) and the columns correspond to the phenomena selected for further study: teams and individual developers. 
The cells are populated with research questions we have brainstormed for further research building in some cases on the ideas we already have, see Table~\ref{tab:AI-step3}. 

\begin{table}[ht]
\centering
\begin{tabular}{p{2cm}|p{5cm}|p{5cm}|}
\cline{2-3}
                               & \multicolumn{2}{|c|}{\textbf{Phenomena to Study (LLMs)}}                            \\ \hline
\multicolumn{1}{|l|}{\textbf{Triadic Dimensions}}  & \textbf{Individual Developers}                                           & \textbf{Team}                                            \\ \hline
\multicolumn{1}{|l|}{\textbf{Enhances}}  & How do LLMs enhance the productivity for developers in terms of writing code?                          & Do LLMs help a team focus more on requirements than on writing code?              \\ \hline
\multicolumn{1}{|l|}{\textbf{Retrieves}} & Do LLMs bring back more focus on using natural language when talking about their code?                                       & How can LLMs be integrated as a Bot or agent as a member of the team? \\ \hline
\multicolumn{1}{|l|}{\textbf{Obsolesces}} & Do LLMs make obsolete some coding skills for developers?         & Do LLMs make the nerd culture obsolete on the team?    \\ \hline
\multicolumn{1}{|l|}{\textbf{Reverses}}  & When used extensively for program synthesis, is there a lack of trust by developers in the code they generate using LLMs? & When used extensively for program synthesis, do LLMs lead to a tangible loss of team knowledge?           \\ \hline
\end{tabular}
\caption{Step 3: Matrix for brainstorming research questions for the disruptive technology of LLMs to Program Synthesis for the phenomena of Individual Developer and Team. These questions are what we the authors brainstormed (as prompted by the playbook).}
\label{tab:AI-step3}
\end{table}



\paragraph*{Step 4: Research method selection and justification}

As a final step, and based on the research questions shown in the matrix created in step 3, the playbook promotes the selection of research strategies that align with the goals of directly studying human participants (e.g., developers or other stakeholders). In Table~\ref{tab:LLM-Step4} we present research strategies aligned with the questions presented in Table~\ref{tab:AI-step3}.

\begin{table}[h]
\centering
\begin{tabular}{p{2cm}|p{5cm}|p{5cm}|}
\cline{2-3}
                               & \multicolumn{2}{|c|}{\textbf{Research Strategies (LLM)}}                            \\ \hline
\multicolumn{1}{|l|}{\textbf{Triadic Dimensions}} & \textbf{Individual Developers} & \textbf{Teams} \\ \hline
\multicolumn{1}{|l|}{\textbf{Enhances}} & Controlled experiments with human participants, large-scale surveys, mining studies? & Field or Case Study, Ethnography, Behavioral measurements\\ \hline
\multicolumn{1}{|l|}{\textbf{Obsolesces}} & Interview and Surveys & Interview and Surveys \\ \hline
\multicolumn{1}{|l|}{\textbf{Retrieves}} & Longitudinal studies  & Field studies \\ \hline
\multicolumn{1}{|l|}{\textbf{Reverses}} & Controlled experiment & Longitudinal case study \\ \hline
\end{tabular}
\caption{Step 4: High-Level Research Strategies to investigate research questions for LLMs on Individual Developers and Teams}
\label{tab:LLM-Step4}
\end{table}

The impact of generative AI on \textbf{enhancing} developer productivity can be assessed through experiments with human participants, as done by Peng et al.~\cite{Peng:copilot:controlledStudy} who investigate the impact of using Copilot on a web-development task. Evidence collected in the lab through controlled studies can be complemented by surveys on the perceived impact of generative AI tools on the developers' productivity. To investigate the impact of LLMs at the team level on requirements engineering, ethnographic field studies might be more suitable. 

As for the questions that consider McLuhan's \textit{retrieve} dimension, a longitudinal observational study may assess how Copilot adoption influences how developers formulate solutions using natural language as they prompt. Indeed, recent research on prompt engineering explores how developers craft natural-language inputs to LLMs to produce their desired outputs. A longitudinal case study can be used to study how bots are adopted and used by a team over time. 

Interview or survey studies could help identify what coding skills are made \textit{obsolete} by the adoption of tools based generative AI and what toxic cultural elements are becoming obsolete because of the emergence of new programming practices supported by these tools at the team level. 

Controlled experiments may align with research questions about the lack of trust by developers in the code generated by humans vs. LLMs (\textit{reverse} dimension). This research approach was successfully adopted to investigate human-bot interaction on Stack Overflow by Murgia et al.~\cite{Murgia:bots:SO}. 
A longitudinal case study, using interviews, may help reveal a potential loss of knowledge at the level of the team over time.

\paragraph*{Reflection on using the Playbook for Research on Generative AI in Program Synthesis}
\label{reflection-playbook-AI}

The playbook is not designed to be prescriptive but rather to aid in an iterative brainstorming process and to help question the choice of research questions and strategies, and along the way to reflect on the impact any potential research questions and designs will lead to.
In this application, the playbook helped identify questions that asked how generative AI could enhance outcomes of program synthesis, but also consider what was made obsolete, what was retrieved and what it may reverse into.  We found it relatively easy to identify what AI may enhance and make obsolete. But what generative AI retrieved took more thought. Ethical issues, that were more of a concern even in the early days of open source and the internet, and chatbots, popular before social media went mainstream, were identified as features retrieved from older technologies.  In terms of what generative AI may reverse into, we found it relatively easy to speculate what may happen if this technology is overused (e.g., lack of trust), but what new technology it will flip into is not yet obvious. 

Our playbook was designed with the goal to promote more consideration of human and social aspects, aspects often overlooked by much of our research ~\cite{storey2020software}. Selecting human and social phenomenon helped steer us in this direction.  But even with a human or social phenomena at the core of a research question, a researcher may choose a research strategy that does not directly involve human participants as the unit of observation.  But this flexibility of how the playbook can be used is a strength, while we hope that using the playbook will help nudge our research in the direction of a heavier emphasis on human and social aspects of software engineering.  

Step 3 involves brainstorming research questions inspired by McLuhan's four laws and social and human phenomena to study.  Identifying research questions  for all cells in the matrix (for the phenomenon we selected) was not trivial. This inspired the development of a customized GPT for our playbook,  based on ChatGPT, to help open up this brainstorming activity. We apply this GPT to the second use case of the research playbook next. Applying the GPT to the above scenario is an exercise we leave for the reader to explore!


\subsection{Applying the Playbook to the Disruption of AR/VR on Software Team Communication and Collaboration}
\label{sec:examples_ARVR}

In recent years, Augmented Reality (AR) and Virtual Reality (VR) technologies have emerged as potent tools with transformative potential in various fields, including software development. Their application in enhancing team communication and collaboration, especially in remote work contexts, has garnered significant attention~\cite{krauss2021current}. Traditional collaboration tools and methods are evolving rapidly, with AR/VR technologies offering immersive and interactive experiences that could redefine the landscape of team dynamics and productivity~\cite{krause2022collaborative}.

We recognize that applying the playbook is not always straightforward (as mentioned above), especially with very new technologies that show potential for disruption in SE but have not been studied much. 
To that end, we designed
``MyResearchPlaybook,'' a specialized GPT model, as a significant adjunct to our research playbook, to help in the brainstorming activity the playbook invokes (see Appendix~\ref{ResearchPlaybookGPT} for details on this GPT model).
Here, we apply MyResearchPlaybook to streamline the process of investigating the complex interactions and impacts of AR/VR technologies on software team collaboration. As a prompt to this specialized GPT model, we asked: \textit{How will AR/VR tools disrupt software team communication and collaboration?} 
 
Below, we show the outcome of using MyResearchPlaybook to assist in this research brainstorming activity (Appendix~\ref{ResearchPlaybookGPT} provides details on how the GPT model was used):

\paragraph*{Step 1: Using McLuhan’s tetrad to conceptualize the influence of AR/VR on team dynamics.}

The application of McLuhan's tetrad by MyResearchPlaybook aids in expanding the scope and depth of research inquiries. We present an iteration of this process as applied to AR/VR on software team communication and collaboration, noting that we manually refined the output generated by MyResearchPlaybook by adding references to literature we found that related to the provided suggestions: 

\begin{itemize}
\item \ul{How will AR/VR \textbf{enhance} communication and collaboration in software teams?} These technologies are likely to introduce more immersive and interactive ways of working together, potentially improving understanding, engagement, and efficiency in collaborative tasks\cite{krause2022collaborative}. Enhanced spatial awareness and visual communication offered by AR/VR can lead to more effective teamwork and problem-solving~\cite{icsmevr23,churchill1998collaborative,ShiSha:2021}.
\item \ul{What traditional methods might AR/VR make \textbf{obsolete} in software team environments?} Traditional video conferencing and text-based communication tools may become less relevant, as AR/VR provide more holistic and engaging interaction methods. The need for physical presence in office spaces could also diminish, as AR/VR create virtual spaces that mimic real-world interaction~\cite{Perry:2016}.
\item \ul{What might AR/VR \textbf{retrieve} that was previously obsolete in team collaboration?} AR/VR could bring back a sense of 'physical' presence and closeness that remote teams have lost, revitalizing the face-to-face interaction dynamics in a digital format~\cite{Fang:CSCW21}.
\item \ul{What could be the \textbf{reversal} effects of extreme use of AR/VR in software teams?} Over-reliance on these technologies could potentially lead to a decrease in face-to-face social skills and physical team bonding. The high immersion could also lead to difficulties in distinguishing between virtual and real-life interactions, and there might be an increased risk of digital fatigue or disorientation~\cite{ar-induce-sickness:2022}.
\end{itemize}

\paragraph*{Step 2: Deciding which phenomena to study and identifying which ideas we may have that we can build on (if any).}

Our playbook emphasizes the importance of human and social dynamics, particularly in the realm of emerging AR/VR technologies. In applying the playbook to this domain, for this step we delve into the multifaceted effects of AR/VR on both individual contributors and collaborative teams. This exploration entails a thorough review of existing research, while also paying close attention to industry trends and real-world applications of AR/VR in software development environments. It is crucial to understand not only the technological capabilities of AR/VR tools but also their nuanced impact on team dynamics, communication styles, and the overall work culture. By incorporating these insights, we aim to offer a comprehensive perspective on how AR/VR technologies are reshaping the landscape of software team collaboration, highlighting both the opportunities and challenges presented by this disruptive innovation.

\paragraph*{Step 3: Creating a matrix of possible research questions concerning phenomena and ideas that can be studied.}

Researchers often commit prematurely to research questions -- this step of the playbook slows down that process and encourages a more careful consideration of the impacts, inspired by the tetrad, crossed with possible phenomena and ideas about those phenomena to brainstorm research questions. As mentioned above, here the playbook encourages us to select and justify specific research questions for further study. For this scenario and step, MyResearchPlaybook generated these questions for us which where each cell contains a potential research question to study. The rows of this matrix correspond to the McLuhan laws (the tetradic dimensions, as mentioned above) and the columns correspond to the phenomena we selected in Step 2 for further study: AR/VR in individual work and AR/VR in team collaboration, 
see Table~\ref{tab:ARVR-step3}.

\begin{table}[ht]
\centering
\begin{tabular}{p{2cm}|p{5cm}|p{5cm}|}
\cline{2-3}
                               & \multicolumn{2}{|c|}{\textbf{Phenomena to Study (AR/VR)}}                            \\ \hline
\multicolumn{1}{|l|}{\textbf{Triadic Dimensions}}  & \textbf{Individual Work}                                           & \textbf{Team Collaboration}                                            \\ \hline
\multicolumn{1}{|l|}{\textbf{Enhances}}  & How does AR/VR enhance individual productivity and engagement in software development tasks?                          & How does AR/VR improve collaborative problem-solving and project management in software teams?              \\ \hline
\multicolumn{1}{|l|}{\textbf{Retrieves}} & Does AR/VR bring back a sense of physical presence and interaction in individual remote work settings?                                       & In what ways does AR/VR retrieve traditional face-to-face collaboration dynamics in software teams? \\ \hline
\multicolumn{1}{|l|}{\textbf{Obsolesces}} & Does AR/VR render traditional desktop-based software development tools obsolete for individual developers?         & Does the use of AR/VR in team settings make conventional remote communication tools (like video calls) obsolete?    \\ \hline
\multicolumn{1}{|l|}{\textbf{Reverses}}  & When used to an extreme, does AR/VR lead to a disconnection from the physical workspace for individual developers? & Could an overreliance on AR/VR technologies in software teams lead to a loss of interpersonal skills and team cohesion?           \\ \hline
\end{tabular}
\caption{Step 3: Matrix for brainstorming research questions for the disruptive technology of AR/VR in Software Team Communication and Collaboration for the phenomena of Individual Work and Team Collaboration. MyResearchPlaybook generated the research questions shown in the Table's cells.}
\label{tab:ARVR-step3}
\end{table}

\paragraph*{Step 4: Research strategy selection and justification}

As a final step, and based on the research questions outlined in the matrix generated in step 3, we again turn to the MyResearchPlaybook to propose research strategies that align with the goals of directly studying human participants (e.g., software developers and teams) and that  address the questions generated in step 3 (see Table~\ref{tab:ARVR-step4}). The impact of AR/VR on individual developers' productivity and engagement can be assessed through experiments with human participants. These studies can be complemented by surveys on the perceived impact of AR/VR tools on developers' productivity and well-being. To investigate the impact of AR/VR at the team level, field studies might be more appropriate, including case studies and ethnographic approaches to understand the nuanced changes in team dynamics and collaboration. 

As for the \textit{retrieve} dimensions, a longitudinal study could help assess the long-term effects of AR/VR adoption on individual work habits and team collaboration practices. 
Interviews or survey studies could be instrumental in identifying the skills and traditional work practices made \textit{obsolete} by the adoption of AR/VR. 
To investigate potential negative effects or over-reliance on AR/VR technologies (\textit{reverse} dimension), controlled experiments and qualitative studies, such as focus groups or in-depth interviews, could be used to gather insights from both individual developers and teams.

\begin{table}[h]
\centering
\begin{tabular}{p{2cm}|p{5cm}|p{5cm}|}
\cline{2-3}
                               & \multicolumn{2}{|c|}{\textbf{Research Strategies (AR/VR)}}                            \\ \hline
\multicolumn{1}{|l|}{\textbf{Triadic Dimensions}} & \textbf{Individual Work} & \textbf{Team Collaboration} \\ \hline
\multicolumn{1}{|l|}{\textbf{Enhances}} & Experiments with human participants, Surveys & Field Studies, Case Studies, Ethnography \\ \hline
\multicolumn{1}{|l|}{\textbf{Obsolesces}} & Interviews, Surveys & Interviews, Surveys \\ \hline
\multicolumn{1}{|l|}{\textbf{Retrieves}} & Longitudinal studies & Longitudinal studies, Field studies \\ \hline
\multicolumn{1}{|l|}{\textbf{Reverses}} & Controlled experiments, Qualitative studies & Qualitative studies, Longitudinal case studies \\ \hline
\end{tabular}
\caption{Step 4: High-Level Research Strategies for AR/VR on Individual Work and Team Collaboration}
\label{tab:ARVR-step4}
\end{table}

\paragraph*{Reflection on using the Playbook for Research on AR/VR on Software Team Communication}
\label{reflection-playbook-ARVR}

Although the MyResearchPlaybook GPT model provided a structured and systematic approach for applying the playbook to the impact of AR/VR on team communication and collaboration, we also found some limitations.  Its effectiveness is reliant on the existing body of research and data available. In the rapidly evolving field of AR/VR, this can be a significant limitation, as current research may not yet fully capture the long-term effects or emerging trends. This limitation potentially lead to gaps in the playbook's ability to generate comprehensive and forward-looking research questions for this use case. Further, the generalized framework provided by MyResearchPlaybook may not adequately address the unique and specific contexts within which AR/VR technologies are deployed. The nuances of different software development environments, team cultures, and individual differences in technology adoption are challenging to encapsulate in a standardized model, possibly leading to oversimplified conclusions. Further iterations to refine the questions need to take these different contexts into consideration, but we see how using the playbook helped with an initial version of research questions.  

Another limitation we note is that the MyResearchPlaybook may operate on the assumption of linear and predictable technological progression. However, the development and adoption of AR/VR technologies are influenced by a myriad of factors, including economic, social, and regulatory aspects, which may lead to non-linear and unforeseen outcomes. This can limit the playbook's ability to accurately forecast and analyze the future implications of AR/VR in software teams. However, this limitation is also something a human may encounter in any case.

We noted, that the playbook, while a powerful tool, required significant human oversight and interpretation. The selection of phenomena, dimensions, and research questions, although guided by MyResearchPlaybook, ultimately depends on the researcher's insights and decisions. The GPT also did not identify literature references for us to consider. Thus, we followed the suggestions with a brief study of the literature to identify references that can inspire the research brainstorming activity. These manual steps introduce a subjective element, which can lead to biases or oversight of critical aspects but are also an asset by keeping human creativity an essential part of our research process.

\section{Provoking a Paradigm Shift in Software Engineering Research} 
\label{sec:discussion}

Technological advancements in software engineering are undeniably shaping the future of the discipline. However, as we discuss the ramifications of disruptive technologies, it is paramount to understand that software engineering is fundamentally a socio-technical domain. Beyond the algorithms and architectures, the human and social dimensions provide a comprehensive perspective on the implications of these technologies. Large-scale software projects, collaborative development practices, and the rise of social coding platforms have transformed not only the methodologies but also the dynamics of who contributes and how they collaborate.

\textbf{McLuhan's Insight on Technologies as Human Extensions}:
McLuhan posited that every technology, or ``media", serves as an extension of human capabilities. Introducing a disruptive technology into software engineering is not just about adding a new tool or methodology. When we introduce disruptive technologies like AR, VR, or AI into the software engineering landscape, we are not merely integrating a new tool~\cite{hicks_psychological_2024}. We are ushering in entities that interact with, augment, and sometimes challenge human capabilities. For instance, the rise of AR and VR is not just about enhanced visualization; it is about redefining how we interact with software systems. Similarly, AI's role is not confined to task automation; it has profound implications on developer roles, team dynamics, and the entire software development lifecycle.

\textbf{Beyond Incrementalism}:
Generative AI, coupled with the immersive experiences offered by AR and VR, is more than a technological leap; it's a paradigm shift. These technologies are not just about enhancing existing practices but are inaugurating a new era of creativity and interaction. We must move beyond the safety of incremental `delta papers' and dare to engage with research that is truly disruptive. This means challenging the traditional benchmarks for evaluating tools and outcomes, urging a broader, more profound, and more meaningful exploration.

\textbf{Redefining Evaluation and Impact}:
Rejecting benchmarks as the sole measure of success, we must embrace a more holistic evaluation of technological impacts. Benchmarks, while valuable, often do not capture the multifaceted impact of disruptive technologies on the socio-technical aspects of software development. A more comprehensive evaluation framework that considers user experience, societal impact, and ethical considerations is critical especially when disruptive technologies are introduced - what matters and what should be valued has possibly changed.

\textbf{Balancing Automation with Augmentation}:
Our focus must extend beyond mere automation. The true potential of disruptive technologies lies in their ability to augment human capabilities, enriching the creative and collaborative aspects of software engineering. This shift from automation to augmentation involves considering how these technologies can enhance, rather than replace, human skills and ingenuity.

\textbf{McGrath's Framework and the Socio-Technical Imperative}:
Building on McGrath's research framework, it is evident that the phenomena we study in software engineering are not just technical constructs. They encompass developer communities, collaboration patterns, ethical considerations, user experiences, and societal impacts. The human and social dimensions, especially in the face of disruptive technologies like AR, VR, and AI, are not mere footnotes. They form the crux of software engineering research. The human and social facets are not peripheral; they are at the heart of software engineering research, especially in the context of disruptive technologies.

\textbf{Framing the Right Questions}:
The trajectory of our research and innovations in software engineering is heavily influenced by the questions we pose. Inspired by McLuhan's tetrad, our inquiries should not be limited to the immediate functionalities or obsolescence of a technology. We must probe deeper into the broader socio-technical ripples. For instance, how does the integration of AI in software engineering reshape team hierarchies or inclusivity paradigms? What emerging practices or norms are developer communities adopting in response to these innovations? How do these technological shifts impact perceptions of software reliability, trust, or ethical standards?
Furthermore, as technologies like generative AI or AR/VR redefine software development's landscape, they challenge established notions of expertise and collaboration. The democratization brought about by these technologies introduces a plethora of voices and perspectives into software engineering. While this diversity is a strength, it also presents challenges in communication, collaboration, and conflict mediation. 
As we find ourselves at the forefront of these technological breakthroughs, it is crucial to pause and reflect on the wider socio-technical consequences. We need to push our boundaries, daring to ask deeper questions that delve into the complex interplay between humans, technologies, and society.

\textbf{Designing the Future of Research Creativity}:
To truly liberate our passion for research in this rapidly evolving landscape, we need to establish structures that foster creativity, interdisciplinary collaboration, and bold exploration. This involves creating academic and industrial environments that support risk-taking, encourage exploration of unconventional ideas, and facilitate the integration of diverse perspectives.

\textbf{Disrupting Research with Disruptive Innovations}: As we explore disruptive innovations that are impacting software engineering, we are also recognizing how these innovations also impact the very essence of how we do research.  The internet and social media both impacted our ways of doing and sharing research, and likewise, tools like ChatGPT are having a similar if not bigger disruption.  We found ourselves using ChatGPT throughout the writing of our paper and design of the playbook (somewhat reluctantly at first), but finally embraced the role ChatGPT can play by designing a custom GPT to support the use of the playbook.  As mentioned above, our intent is that it should not replace the creativity and deep reflection we as researchers do, but it can be used to support brainstorming and pointers to ideas and resources we may not have known about. 

In summary, as we navigate the complexities of disruptive technologies in software engineering, our role as researchers and practitioners demands a provocative, forward-thinking approach. We must challenge existing paradigms, embrace the socio-technical nature of our field, and seek to understand and shape the broader implications of our work. By doing so, we can ensure that our contributions to the field of software engineering are not only technologically advanced but also socially responsible and ethically sound. It is through this holistic understanding and approach that we can truly harness the transformative power of disruptive technologies in software engineering.

\section{Conclusion}

Software engineering, as a discipline, has seen, since its inception, many disruptive technologies that have augmented and extended the capabilities of its engineers who design, write, maintain and ensure the quality of software programs. These disruptions, like other forms of media and technologies, all follow the same four laws that McLuhan noted in that they will enhance software engineering capabilities, they will make obsolete some existing technologies, they will retrieve aspects of prior technologies, and they will ultimately reverse the desired outcomes that inspired them when used over time. 

 As researchers, we have the privilege and opportunity to study the impacts these technologies introduce. However, often our research is slow to respond to the disruptions new technologies evoke, and we take an overly narrow approach and focus on the short term technical advantages they introduce. McLuhan recognized this tendency for people to focus on positive influences decades ago and his observations of the four laws of media led to his tetrad of provoking and non-obvious questions to ask about new technologies. 
 
 The playbook we introduce, builds on McLuhan's four laws and likewise is designed to provoke and prompt new research directions that ask more critical questions reflecting on what has come before as well as what will happen, especially to socio-technical aspects of software engineering, when the technologies are used and possibly overused over time.  We combine McLuhan's tetrad with research frameworks that help us craft pertinent research questions about socio-technical aspects in software engineering, while also selecting research strategies that are aligned with studying the phenomena that may matter the most over time - that of human and social aspects. 

 We apply the playbook in two ways. First we use it to reflect on a prior disruptive technology on software engineering---Stack Overflow---and to consider some of the landscape of research that investigated the impact of it on individual developers and development teams. This retrospective application of the playbook not only illustrated the playbook, but also revealed that this prior research was predominantly focused on McLuhan's ``enhance'' and ``obsolete'' dimensions, and this research had, especially initially, little emphasis on studying social and human aspects.  We also applied the playbook to two disruptive technologies that are showing disruption on software engineering.  The playbook helps steer towards new research directions that focus on human and social aspects.  Finally, along with the playbook, we provide a customized ChatGPT called MyResearchPlaybook that some researchers may find useful as they brainstorm and design their research questions and studies. We expect we are just at the beginning of these new research assistant tools that will help us expand, accelerate, broaden and improve our research over time. 
 
 In conclusion, software is designed by people for people. How technology will change human lives, capabilities and values changes faster than we can imagine, and demands that we anticipate and pay close attention to the changes that may occur.  The playbook we present, we hope, will provide a way for researchers to pause before selecting a particular research question or strategy, and to reflect on what else they could study that may be ultimately more important to consider with such an uncertain future in mind.   

\begin{acks}
We are grateful to the organizers and participants of the NII Shonan Meeting (No.191) ``Human Aspects in Software Engineering" (March 2023) for their valuable insights and discussions that inspired the genesis of this playbook. Additionally, our thanks go to the attendees of The Copenhagen Symposium on Human-Centered Software Engineering AI (November 2023) whose constructive feedback was crucial in refining our work (Grants G-2023-21020 and CF23-0208). The contributions from both events have been instrumental in shaping the perspectives presented in this study. We also thank Cassandra Petrachenko for her carefully editing and improving our paper. 
\end{acks}

\bibliographystyle{ACM-Reference-Format}
\bibliography{references}


\begin{thebibliography}{89}


\ifx \showCODEN    \undefined \def \showCODEN     #1{\unskip}     \fi
\ifx \showDOI      \undefined \def \showDOI       #1{#1}\fi
\ifx \showISBNx    \undefined \def \showISBNx     #1{\unskip}     \fi
\ifx \showISBNxiii \undefined \def \showISBNxiii  #1{\unskip}     \fi
\ifx \showISSN     \undefined \def \showISSN      #1{\unskip}     \fi
\ifx \showLCCN     \undefined \def \showLCCN      #1{\unskip}     \fi
\ifx \shownote     \undefined \def \shownote      #1{#1}          \fi
\ifx \showarticletitle \undefined \def \showarticletitle #1{#1}   \fi
\ifx \showURL      \undefined \def \showURL       {\relax}        \fi
\providecommand\bibfield[2]{#2}
\providecommand\bibinfo[2]{#2}
\providecommand\natexlab[1]{#1}
\providecommand\showeprint[2][]{arXiv:#2}

\bibitem[cop(2021)]%
        {copilot2021github}
 \bibinfo{year}{2021}\natexlab{}.
\newblock \bibinfo{title}{GitHub Copilot}.
\newblock
\newblock
\urldef\tempurl%
\url{https://copilot.github.com}
\showURL{%
\tempurl}


\bibitem[Abdalkareem et~al\mbox{.}(2017)]%
        {ABDALKAREEM2017148}
\bibfield{author}{\bibinfo{person}{Rabe Abdalkareem}, \bibinfo{person}{Emad
  Shihab}, {and} \bibinfo{person}{Juergen Rilling}.}
  \bibinfo{year}{2017}\natexlab{}.
\newblock \showarticletitle{On code reuse from StackOverflow: An exploratory
  study on Android apps}.
\newblock \bibinfo{journal}{\emph{Information and Software Technology}}
  \bibinfo{volume}{88} (\bibinfo{year}{2017}), \bibinfo{pages}{148--158}.
\newblock
\showISSN{0950-5849}
\urldef\tempurl%
\url{https://doi.org/10.1016/j.infsof.2017.04.005}
\showDOI{\tempurl}


\bibitem[Acar et~al\mbox{.}(2016a)]%
        {Yasemin:security:2016}
\bibfield{author}{\bibinfo{person}{Yasemin Acar}, \bibinfo{person}{Michael
  Backes}, \bibinfo{person}{Sascha Fahl}, \bibinfo{person}{Doowon Kim},
  \bibinfo{person}{Michelle~L. Mazurek}, {and} \bibinfo{person}{Christian
  Stransky}.} \bibinfo{year}{2016}\natexlab{a}.
\newblock \showarticletitle{You Get Where You're Looking for: The Impact of
  Information Sources on Code Security}. In \bibinfo{booktitle}{\emph{2016 IEEE
  Symposium on Security and Privacy (SP)}}. \bibinfo{pages}{289--305}.
\newblock
\urldef\tempurl%
\url{https://doi.org/10.1109/SP.2016.25}
\showDOI{\tempurl}


\bibitem[Acar et~al\mbox{.}(2016b)]%
        {Acar:SO:codeSecurity:2016}
\bibfield{author}{\bibinfo{person}{Yasemin Acar}, \bibinfo{person}{Michael
  Backes}, \bibinfo{person}{Sascha Fahl}, \bibinfo{person}{Doowon Kim},
  \bibinfo{person}{Michelle~L. Mazurek}, {and} \bibinfo{person}{Christian
  Stransky}.} \bibinfo{year}{2016}\natexlab{b}.
\newblock \showarticletitle{You Get Where You're Looking for: The Impact of
  Information Sources on Code Security}. In \bibinfo{booktitle}{\emph{2016 IEEE
  Symposium on Security and Privacy (SP)}}. \bibinfo{pages}{289--305}.
\newblock
\urldef\tempurl%
\url{https://doi.org/10.1109/SP.2016.25}
\showDOI{\tempurl}


\bibitem[Anderson et~al\mbox{.}(2013)]%
        {anderson2013badges}
\bibfield{author}{\bibinfo{person}{Ashton Anderson}, \bibinfo{person}{Daniel
  Huttenlocher}, \bibinfo{person}{Jon Kleinberg}, {and} \bibinfo{person}{Jure
  Leskovec}.} \bibinfo{year}{2013}\natexlab{}.
\newblock \showarticletitle{Steering User Behavior with Badges}. In
  \bibinfo{booktitle}{\emph{Proceedings of the 22nd International Conference on
  World Wide Web}} (Rio de Janeiro, Brazil) \emph{(\bibinfo{series}{WWW '13})}.
  \bibinfo{publisher}{Association for Computing Machinery},
  \bibinfo{address}{New York, NY, USA}, \bibinfo{pages}{95–106}.
\newblock
\showISBNx{9781450320351}
\urldef\tempurl%
\url{https://doi.org/10.1145/2488388.2488398}
\showDOI{\tempurl}


\bibitem[Asaduzzaman et~al\mbox{.}(2013)]%
        {Asaduzzaman2013Answering}
\bibfield{author}{\bibinfo{person}{Muhammad Asaduzzaman},
  \bibinfo{person}{Ahmed~Shah Mashiyat}, \bibinfo{person}{Chanchal~K Roy},
  {and} \bibinfo{person}{Kevin~A Schneider}.} \bibinfo{year}{2013}\natexlab{}.
\newblock \showarticletitle{Answering questions about unanswered questions of
  {Stack Overflow}}. In \bibinfo{booktitle}{\emph{Proceedings of the 10th
  International Working Conference on Mining Software Repositories}}.
  \bibinfo{publisher}{IEEE}, \bibinfo{pages}{97--100}.
\newblock


\bibitem[Bacchelli et~al\mbox{.}(2012)]%
        {Bacchelli:etAl:IDE}
\bibfield{author}{\bibinfo{person}{Alberto Bacchelli}, \bibinfo{person}{Luca
  Ponzanelli}, {and} \bibinfo{person}{Michele Lanza}.}
  \bibinfo{year}{2012}\natexlab{}.
\newblock \showarticletitle{Harnessing Stack Overflow for the IDE}. In
  \bibinfo{booktitle}{\emph{2012 Third International Workshop on Recommendation
  Systems for Software Engineering (RSSE)}}. \bibinfo{pages}{26--30}.
\newblock
\urldef\tempurl%
\url{https://doi.org/10.1109/RSSE.2012.6233404}
\showDOI{\tempurl}


\bibitem[Barenkamp et~al\mbox{.}(2020)]%
        {barenkamp2020applications}
\bibfield{author}{\bibinfo{person}{Marco Barenkamp}, \bibinfo{person}{Jonas
  Rebstadt}, {and} \bibinfo{person}{Oliver Thomas}.}
  \bibinfo{year}{2020}\natexlab{}.
\newblock \showarticletitle{Applications of AI in classical software
  engineering}.
\newblock \bibinfo{journal}{\emph{AI Perspectives}} \bibinfo{volume}{2},
  \bibinfo{number}{1} (\bibinfo{year}{2020}), \bibinfo{pages}{1}.
\newblock


\bibitem[Barua et~al\mbox{.}(2014a)]%
        {Barua:etAl:SOTopics}
\bibfield{author}{\bibinfo{person}{Anton Barua}, \bibinfo{person}{Stephen~W.
  Thomas}, {and} \bibinfo{person}{Ahmed~E. Hassan}.}
  \bibinfo{year}{2014}\natexlab{a}.
\newblock \showarticletitle{What Are Developers Talking about? An Analysis of
  Topics and Trends in Stack Overflow}.
\newblock \bibinfo{journal}{\emph{Empirical Softw. Engg.}}
  \bibinfo{volume}{19}, \bibinfo{number}{3} (\bibinfo{date}{jun}
  \bibinfo{year}{2014}), \bibinfo{pages}{619–654}.
\newblock
\showISSN{1382-3256}
\urldef\tempurl%
\url{https://doi.org/10.1007/s10664-012-9231-y}
\showDOI{\tempurl}


\bibitem[Barua et~al\mbox{.}(2014b)]%
        {barua2014developers}
\bibfield{author}{\bibinfo{person}{Anton Barua}, \bibinfo{person}{Stephen~W
  Thomas}, {and} \bibinfo{person}{Ahmed~E Hassan}.}
  \bibinfo{year}{2014}\natexlab{b}.
\newblock \showarticletitle{What are developers talking about? an analysis of
  topics and trends in stack overflow}.
\newblock \bibinfo{journal}{\emph{Empirical software engineering}}
  \bibinfo{volume}{19} (\bibinfo{year}{2014}), \bibinfo{pages}{619--654}.
\newblock


\bibitem[Barzilay et~al\mbox{.}(2013)]%
        {barzilay2013facilitating}
\bibfield{author}{\bibinfo{person}{Ohad Barzilay}, \bibinfo{person}{Christoph
  Treude}, {and} \bibinfo{person}{Alexey Zagalsky}.}
  \bibinfo{year}{2013}\natexlab{}.
\newblock \showarticletitle{Facilitating crowd sourced software engineering via
  stack overflow}.
\newblock \bibinfo{journal}{\emph{Finding Source Code on the Web for Remix and
  Reuse}} (\bibinfo{year}{2013}), \bibinfo{pages}{289--308}.
\newblock


\bibitem[Bernstein and Noy(2014)]%
        {bernstein14}
\bibfield{author}{\bibinfo{person}{Abraham Bernstein} {and}
  \bibinfo{person}{Natasha Noy}.} \bibinfo{year}{2014}\natexlab{}.
\newblock \bibinfo{booktitle}{\emph{Is This Really Science? The Semantic
  Webber’s Guide to Evaluating Research Contributions}}.
\newblock \bibinfo{type}{{T}echnical {R}eport}.
\newblock


\bibitem[Bhasin et~al\mbox{.}(2021)]%
        {bhasin2021student}
\bibfield{author}{\bibinfo{person}{Trishala Bhasin}, \bibinfo{person}{Adam
  Murray}, {and} \bibinfo{person}{Margaret-Anne Storey}.}
  \bibinfo{year}{2021}\natexlab{}.
\newblock \showarticletitle{Student experiences with github and stack overflow:
  An exploratory study}. In \bibinfo{booktitle}{\emph{2021 IEEE/ACM 13th
  International Workshop on Cooperative and Human Aspects of Software
  Engineering (CHASE)}}. IEEE, \bibinfo{pages}{81--90}.
\newblock


\bibitem[Bosu et~al\mbox{.}(2013)]%
        {Bosu2013Building}
\bibfield{author}{\bibinfo{person}{Amiangshu Bosu},
  \bibinfo{person}{Christopher~S Corley}, \bibinfo{person}{Dustin Heaton},
  \bibinfo{person}{Debarshi Chatterji}, \bibinfo{person}{Jeffrey~C Carver},
  {and} \bibinfo{person}{Nicholas~A Kraft}.} \bibinfo{year}{2013}\natexlab{}.
\newblock \showarticletitle{Building reputation in {StackOverflow}: {A}n
  empirical investigation}. In \bibinfo{booktitle}{\emph{Proceedings of the
  10th International Working Conference on Mining Software Repositories}}.
  \bibinfo{publisher}{IEEE}, \bibinfo{pages}{89--92}.
\newblock


\bibitem[Bower and Christensen(1995)]%
        {christensen:95}
\bibfield{author}{\bibinfo{person}{Joseph~L. Bower} {and}
  \bibinfo{person}{Clayton~M. Christensen}.} \bibinfo{year}{1995}\natexlab{}.
\newblock \showarticletitle{Disruptive Technologies: Catching the Wave}.
\newblock In \bibinfo{booktitle}{\emph{Harvard Business Review}}.
  \bibinfo{pages}{43--53}.
\newblock
Issue January--February 1995.


\bibitem[Burtch et~al\mbox{.}(2023)]%
        {Burtch:GPTvsSO:2023}
\bibfield{author}{\bibinfo{person}{Gordon Burtch}, \bibinfo{person}{Dokyun
  Lee}, {and} \bibinfo{person}{Zhichen Chen}.} \bibinfo{year}{2023}\natexlab{}.
\newblock \showarticletitle{The Consequences of Generative AI for UGC and
  Online Community Engagement}.
\newblock \bibinfo{journal}{\emph{IEEE Transactions on Software Engineering}}
  (\bibinfo{year}{2023}).
\newblock
\urldef\tempurl%
\url{https://doi.org/10.2139/ssrn.4521754}
\showDOI{\tempurl}


\bibitem[Calefato et~al\mbox{.}(2015)]%
        {Calefato:successfulAnswers:2015}
\bibfield{author}{\bibinfo{person}{Fabio Calefato}, \bibinfo{person}{Filippo
  Lanubile}, \bibinfo{person}{Maria~Concetta Marasciulo}, {and}
  \bibinfo{person}{Nicole Novielli}.} \bibinfo{year}{2015}\natexlab{}.
\newblock \showarticletitle{Mining Successful Answers in Stack Overflow}. In
  \bibinfo{booktitle}{\emph{2015 IEEE/ACM 12th Working Conference on Mining
  Software Repositories}}. \bibinfo{pages}{430--433}.
\newblock
\urldef\tempurl%
\url{https://doi.org/10.1109/MSR.2015.56}
\showDOI{\tempurl}


\bibitem[Calefato et~al\mbox{.}(2018)]%
        {Calefato:etAl:2018:howtoask}
\bibfield{author}{\bibinfo{person}{Fabio Calefato}, \bibinfo{person}{Filippo
  Lanubile}, {and} \bibinfo{person}{Nicole Novielli}.}
  \bibinfo{year}{2018}\natexlab{}.
\newblock \showarticletitle{How to Ask for Technical Help? Evidence-Based
  Guidelines for Writing Questions on Stack Overflow}.
\newblock \bibinfo{journal}{\emph{Inf. Softw. Technol.}} \bibinfo{volume}{94},
  \bibinfo{number}{C} (\bibinfo{date}{feb} \bibinfo{year}{2018}),
  \bibinfo{pages}{186–207}.
\newblock
\showISSN{0950-5849}


\bibitem[Campos et~al\mbox{.}(2016)]%
        {CAMPOS:2016-bugFixing}
\bibfield{author}{\bibinfo{person}{Eduardo~C. Campos}, \bibinfo{person}{Martin
  Monperrus}, {and} \bibinfo{person}{Marcelo~A. Maia}.}
  \bibinfo{year}{2016}\natexlab{}.
\newblock \showarticletitle{Searching Stack Overflow for API-Usage-Related Bug
  Fixes Using Snippet-Based Queries} \emph{(\bibinfo{series}{CASCON '16})}.
  \bibinfo{publisher}{IBM Corp.}, \bibinfo{address}{USA},
  \bibinfo{pages}{232–242}.
\newblock


\bibitem[Cao et~al\mbox{.}(2021)]%
        {Cao:etAl:EfficientSearch}
\bibfield{author}{\bibinfo{person}{Kaibo Cao}, \bibinfo{person}{Chunyang Chen},
  \bibinfo{person}{Sebastian Baltes}, \bibinfo{person}{Christoph Treude}, {and}
  \bibinfo{person}{Xiang Chen}.} \bibinfo{year}{2021}\natexlab{}.
\newblock \showarticletitle{Automated Query Reformulation for Efficient Search
  Based on Query Logs From Stack Overflow}. In \bibinfo{booktitle}{\emph{2021
  IEEE/ACM 43rd International Conference on Software Engineering (ICSE)}}.
  \bibinfo{pages}{1273--1285}.
\newblock
\urldef\tempurl%
\url{https://doi.org/10.1109/ICSE43902.2021.00116}
\showDOI{\tempurl}


\bibitem[Chatterjee et~al\mbox{.}(2020)]%
        {CHATTERJEE2020110454}
\bibfield{author}{\bibinfo{person}{Preetha Chatterjee}, \bibinfo{person}{Minji
  Kong}, {and} \bibinfo{person}{Lori Pollock}.}
  \bibinfo{year}{2020}\natexlab{}.
\newblock \showarticletitle{Finding help with programming errors: An
  exploratory study of novice software engineers’ focus in stack overflow
  posts}.
\newblock \bibinfo{journal}{\emph{Journal of Systems and Software}}
  \bibinfo{volume}{159} (\bibinfo{year}{2020}), \bibinfo{pages}{110454}.
\newblock
\showISSN{0164-1212}
\urldef\tempurl%
\url{https://doi.org/10.1016/j.jss.2019.110454}
\showDOI{\tempurl}


\bibitem[Chen et~al\mbox{.}(2021)]%
        {chen2021evaluating}
\bibfield{author}{\bibinfo{person}{Mark Chen}, \bibinfo{person}{Jakub Tworek},
  \bibinfo{person}{Heewoo Jun}, \bibinfo{person}{Qijing Yuan},
  \bibinfo{person}{Henryk P. de~Oliveira Pinto}, \bibinfo{person}{Jared
  Kaplan}, \bibinfo{person}{Harrison Edwards}, \bibinfo{person}{Yuri Burda},
  \bibinfo{person}{Niru Joseph}, \bibinfo{person}{Greg Brockman},
  \bibinfo{person}{Alexander Ray}, \bibinfo{person}{Rishita Puri},
  \bibinfo{person}{Gabriel Krueger}, \bibinfo{person}{Mike Petrov},
  \bibinfo{person}{Hala Khlaaf}, \bibinfo{person}{Girish Sastry},
  \bibinfo{person}{Pamela Mishkin}, \bibinfo{person}{Ben Chan},
  \bibinfo{person}{Scott Gray}, \bibinfo{person}{Nick Ryder},
  \bibinfo{person}{Maxim Pavlov}, \bibinfo{person}{Alex Power},
  \bibinfo{person}{Lukasz Kaiser}, \bibinfo{person}{Maximilian Bavarian},
  \bibinfo{person}{Carolyn Winter}, \bibinfo{person}{Philippe Tillet},
  \bibinfo{person}{Felipe~P. Such}, \bibinfo{person}{David Cummings},
  \bibinfo{person}{Matthias Plappert}, \bibinfo{person}{Fotios Chantzis},
  \bibinfo{person}{Eric Barnes}, \bibinfo{person}{Ariel Herbert-Voss},
  \bibinfo{person}{William~H. Guss}, \bibinfo{person}{Alex Nichol},
  \bibinfo{person}{Andrew Paino}, \bibinfo{person}{Nick Tezak},
  \bibinfo{person}{Jerry Tang}, \bibinfo{person}{Igor Babuschkin},
  \bibinfo{person}{Sujith Balaji}, \bibinfo{person}{Shivendra Jain},
  \bibinfo{person}{Will Saunders}, \bibinfo{person}{Christopher Hesse},
  \bibinfo{person}{Andrew~N. Carr}, \bibinfo{person}{Jan Leike},
  \bibinfo{person}{Joshua Achiam}, \bibinfo{person}{Vedant Misra},
  \bibinfo{person}{Eri Morikawa}, \bibinfo{person}{Alec Radford},
  \bibinfo{person}{Melody Knight}, \bibinfo{person}{Miles Brundage},
  \bibinfo{person}{Matej Murati}, \bibinfo{person}{Katja Mayer},
  \bibinfo{person}{Peter Welinder}, \bibinfo{person}{Brian McGrew},
  \bibinfo{person}{Dario Amodei}, \bibinfo{person}{Sam McCandlish},
  \bibinfo{person}{Ilya Sutskever}, {and} \bibinfo{person}{Wojciech Zaremba}.}
  \bibinfo{year}{2021}\natexlab{}.
\newblock \bibinfo{title}{Evaluating large language models trained on code}.
\newblock
\newblock
\urldef\tempurl%
\url{https://arxiv.org/abs/2107.03374}
\showURL{%
\tempurl}


\bibitem[Christensen and Overdorf(2000)]%
        {christensen:00}
\bibfield{author}{\bibinfo{person}{Clayton~M. Christensen} {and}
  \bibinfo{person}{Michael Overdorf}.} \bibinfo{year}{2000}\natexlab{}.
\newblock \showarticletitle{Meeting the Challenge of Disruptive Change}.
\newblock In \bibinfo{booktitle}{\emph{Harvard Business Review}}.
  \bibinfo{pages}{66--76}.
\newblock
Issue March--April 2000.


\bibitem[Churchill and Snowdon(1998)]%
        {churchill1998collaborative}
\bibfield{author}{\bibinfo{person}{Elizabeth~F Churchill} {and}
  \bibinfo{person}{Dave Snowdon}.} \bibinfo{year}{1998}\natexlab{}.
\newblock \showarticletitle{Collaborative virtual environments: an introductory
  review of issues and systems}.
\newblock \bibinfo{journal}{\emph{virtual reality}}  \bibinfo{volume}{3}
  (\bibinfo{year}{1998}), \bibinfo{pages}{3--15}.
\newblock


\bibitem[Digkas et~al\mbox{.}(2019)]%
        {Digkas:etAl:2019:reusingSO}
\bibfield{author}{\bibinfo{person}{Georgios Digkas}, \bibinfo{person}{Nikolaos
  Nikolaidis}, \bibinfo{person}{Apostolos Ampatzoglou}, {and}
  \bibinfo{person}{Alexander Chatzigeorgiou}.} \bibinfo{year}{2019}\natexlab{}.
\newblock \showarticletitle{Reusing Code from StackOverflow: The Effect on
  Technical Debt}. In \bibinfo{booktitle}{\emph{2019 45th Euromicro Conference
  on Software Engineering and Advanced Applications (SEAA)}}.
  \bibinfo{pages}{87--91}.
\newblock
\urldef\tempurl%
\url{https://doi.org/10.1109/SEAA.2019.00022}
\showDOI{\tempurl}


\bibitem[Dominic et~al\mbox{.}(2020)]%
        {Dominic2020}
\bibfield{author}{\bibinfo{person}{James Dominic}, \bibinfo{person}{Jada
  Houser}, \bibinfo{person}{Igor Steinmacher}, \bibinfo{person}{Charles
  Ritter}, {and} \bibinfo{person}{Paige Rodeghero}.}
  \bibinfo{year}{2020}\natexlab{}.
\newblock \showarticletitle{Conversational Bot for Newcomers Onboarding to Open
  Source Projects}. In \bibinfo{booktitle}{\emph{Proceedings of the IEEE/ACM
  42nd International Conference on Software Engineering Workshops}} (Seoul,
  Republic of Korea) \emph{(\bibinfo{series}{ICSEW'20})}.
  \bibinfo{publisher}{Association for Computing Machinery},
  \bibinfo{address}{New York, NY, USA}, \bibinfo{pages}{46–50}.
\newblock
\showISBNx{9781450379632}
\urldef\tempurl%
\url{https://doi.org/10.1145/3387940.3391534}
\showDOI{\tempurl}


\bibitem[Easterbrook et~al\mbox{.}(2008)]%
        {easterbrook_selecting_2008}
\bibfield{author}{\bibinfo{person}{Steve Easterbrook}, \bibinfo{person}{Janice
  Singer}, \bibinfo{person}{Margaret-Anne Storey}, {and}
  \bibinfo{person}{Daniela Damian}.} \bibinfo{year}{2008}\natexlab{}.
\newblock \showarticletitle{Selecting {Empirical} {Methods} for {Software}
  {Engineering} {Research}}.
\newblock In \bibinfo{booktitle}{\emph{Guide to {Advanced} {Empirical}
  {Software} {Engineering}}}, \bibfield{editor}{\bibinfo{person}{Forrest
  Shull}, \bibinfo{person}{Janice Singer}, {and} \bibinfo{person}{Dag I.~K.
  Sjoberg}} (Eds.). \bibinfo{publisher}{Springer}, \bibinfo{address}{London},
  \bibinfo{pages}{285--311}.
\newblock
\showISBNx{978-1-84800-044-5}
\urldef\tempurl%
\url{https://doi.org/10.1007/978-1-84800-044-5_11}
\showDOI{\tempurl}


\bibitem[Ebert and Louridas(2023)]%
        {Ebert:Louridas:2023}
\bibfield{author}{\bibinfo{person}{Christof Ebert} {and} \bibinfo{person}{Panos
  Louridas}.} \bibinfo{year}{2023}\natexlab{}.
\newblock \showarticletitle{Generative AI for Software Practitioners}.
\newblock \bibinfo{journal}{\emph{IEEE Software}} \bibinfo{volume}{40},
  \bibinfo{number}{4} (\bibinfo{year}{2023}), \bibinfo{pages}{30--38}.
\newblock
\urldef\tempurl%
\url{https://doi.org/10.1109/MS.2023.3265877}
\showDOI{\tempurl}


\bibitem[Engstr{\"o}m et~al\mbox{.}(2020)]%
        {engstrom2020software}
\bibfield{author}{\bibinfo{person}{Emelie Engstr{\"o}m},
  \bibinfo{person}{Margaret-Anne Storey}, \bibinfo{person}{Per Runeson},
  \bibinfo{person}{Martin H{\"o}st}, {and} \bibinfo{person}{Maria~Teresa
  Baldassarre}.} \bibinfo{year}{2020}\natexlab{}.
\newblock \showarticletitle{How software engineering research aligns with
  design science: a review}.
\newblock \bibinfo{journal}{\emph{Empirical Software Engineering}}
  \bibinfo{volume}{25} (\bibinfo{year}{2020}), \bibinfo{pages}{2630--2660}.
\newblock


\bibitem[Fang et~al\mbox{.}(2021)]%
        {Fang:CSCW21}
\bibfield{author}{\bibinfo{person}{Jingchao Fang}, \bibinfo{person}{Victoria
  Chang}, \bibinfo{person}{Ge Gao}, {and} \bibinfo{person}{Hao-Chuan Wang}.}
  \bibinfo{year}{2021}\natexlab{}.
\newblock \showarticletitle{Social Interactions in Virtual Reality: What Cues
  Do People Use Most and How}. In \bibinfo{booktitle}{\emph{Companion
  Publication of the 2021 Conference on Computer Supported Cooperative Work and
  Social Computing}} (Virtual Event, USA) \emph{(\bibinfo{series}{CSCW '21
  Companion})}. \bibinfo{publisher}{Association for Computing Machinery},
  \bibinfo{address}{New York, NY, USA}, \bibinfo{pages}{49–52}.
\newblock
\showISBNx{9781450384797}
\urldef\tempurl%
\url{https://doi.org/10.1145/3462204.3481772}
\showDOI{\tempurl}


\bibitem[Fernandes and Werner(2022)]%
        {fernandes2022systematic}
\bibfield{author}{\bibinfo{person}{Flipe Fernandes} {and}
  \bibinfo{person}{Cl{\'a}udia Werner}.} \bibinfo{year}{2022}\natexlab{}.
\newblock \showarticletitle{A Systematic Literature Review of the Metaverse for
  Software Engineering Education: Overview, Challenges and Opportunities}.
\newblock \bibinfo{journal}{\emph{PRESENCE: Washington, WA, USA}}
  (\bibinfo{year}{2022}).
\newblock


\bibitem[Fischer et~al\mbox{.}(2017)]%
        {Fischer:security:2017}
\bibfield{author}{\bibinfo{person}{Felix Fischer}, \bibinfo{person}{Konstantin
  Böttinger}, \bibinfo{person}{Huang Xiao}, \bibinfo{person}{Christian
  Stransky}, \bibinfo{person}{Yasemin Acar}, \bibinfo{person}{Michael Backes},
  {and} \bibinfo{person}{Sascha Fahl}.} \bibinfo{year}{2017}\natexlab{}.
\newblock \showarticletitle{Stack Overflow Considered Harmful? The Impact of
  Copy \& Paste on Android Application Security}. In
  \bibinfo{booktitle}{\emph{2017 IEEE Symposium on Security and Privacy (SP)}}.
  \bibinfo{pages}{121--136}.
\newblock
\urldef\tempurl%
\url{https://doi.org/10.1109/SP.2017.31}
\showDOI{\tempurl}


\bibitem[Ford et~al\mbox{.}(2018)]%
        {Ford_etAl:mentoring}
\bibfield{author}{\bibinfo{person}{Denae Ford}, \bibinfo{person}{Kristina
  Lustig}, \bibinfo{person}{Jeremy Banks}, {and} \bibinfo{person}{Chris
  Parnin}.} \bibinfo{year}{2018}\natexlab{}.
\newblock \showarticletitle{"We Don't Do That Here": How Collaborative Editing
  with Mentors Improves Engagement in Social Q\&A Communities}. In
  \bibinfo{booktitle}{\emph{Proceedings of the 2018 CHI Conference on Human
  Factors in Computing Systems}} (Montreal QC, Canada)
  \emph{(\bibinfo{series}{CHI '18})}. \bibinfo{publisher}{Association for
  Computing Machinery}, \bibinfo{address}{New York, NY, USA},
  \bibinfo{pages}{1–12}.
\newblock
\showISBNx{9781450356206}
\urldef\tempurl%
\url{https://doi.org/10.1145/3173574.3174182}
\showDOI{\tempurl}


\bibitem[Ford et~al\mbox{.}(2016)]%
        {ford2016paradise}
\bibfield{author}{\bibinfo{person}{Denae Ford}, \bibinfo{person}{Justin Smith},
  \bibinfo{person}{Philip~J Guo}, {and} \bibinfo{person}{Chris Parnin}.}
  \bibinfo{year}{2016}\natexlab{}.
\newblock \showarticletitle{Paradise unplugged: Identifying barriers for female
  participation on stack overflow}. In \bibinfo{booktitle}{\emph{Proceedings of
  the 2016 24th ACM SIGSOFT International symposium on foundations of software
  engineering}}. \bibinfo{pages}{846--857}.
\newblock


\bibitem[Grant and Betts(2013)]%
        {Grant:Buddy:gamification}
\bibfield{author}{\bibinfo{person}{Scott Grant} {and} \bibinfo{person}{Buddy
  Betts}.} \bibinfo{year}{2013}\natexlab{}.
\newblock \showarticletitle{Encouraging user behaviour with achievements: An
  empirical study}. In \bibinfo{booktitle}{\emph{2013 10th Working Conference
  on Mining Software Repositories (MSR)}}. \bibinfo{pages}{65--68}.
\newblock


\bibitem[Hicks(2024)]%
        {hicks_psychological_2024}
\bibfield{author}{\bibinfo{person}{Catherine~M. Hicks}.}
  \bibinfo{year}{2024}\natexlab{}.
\newblock \showarticletitle{Psychological {Affordances} {Can} {Provide} a
  {Missing} {Explanatory} {Layer} for {Why} {Interventions} to {Improve}
  {Developer} {Experience} {Take} {Hold} or {Fail}}.
\newblock  (\bibinfo{year}{2024}).
\newblock
\urldef\tempurl%
\url{https://files.osf.io/v1/resources/qz43x/providers/osfstorage/65b2f3ae4aa63c07d9df22ec?action=download&direct&version=5}
\showURL{%
\tempurl}
\newblock
\shownote{Publisher: OSF}.


\bibitem[Hoff et~al\mbox{.}(2023)]%
        {icsmevr23}
\bibfield{author}{\bibinfo{person}{Adrian Hoff}, \bibinfo{person}{Christoph
  Seidl}, \bibinfo{person}{Mircea~F Lungu}, {and} \bibinfo{person}{Michele
  Lanza}.} \bibinfo{year}{2023}\natexlab{}.
\newblock \showarticletitle{Preparing Software Re-Engineering via Freehand
  Sketches in Virtual Reality}. In \bibinfo{booktitle}{\emph{39th IEEE
  International Conference on Software Maintenance and Evolution}}. IEEE,
  \bibinfo{pages}{to appear}.
\newblock


\bibitem[Hoppe et~al\mbox{.}(2021)]%
        {ShiSha:2021}
\bibfield{author}{\bibinfo{person}{Adrian~H. Hoppe}, \bibinfo{person}{Florian
  van~de Camp}, {and} \bibinfo{person}{Rainer Stiefelhagen}.}
  \bibinfo{year}{2021}\natexlab{}.
\newblock \showarticletitle{ShiSha: Enabling Shared Perspective With
  Face-to-Face Collaboration Using Redirected Avatars in Virtual Reality}.
\newblock \bibinfo{journal}{\emph{Proc. ACM Hum.-Comput. Interact.}}
  \bibinfo{volume}{4}, \bibinfo{number}{CSCW3}, Article
  \bibinfo{articleno}{251} (\bibinfo{date}{jan} \bibinfo{year}{2021}),
  \bibinfo{numpages}{22}~pages.
\newblock
\urldef\tempurl%
\url{https://doi.org/10.1145/3432950}
\showDOI{\tempurl}


\bibitem[Hou et~al\mbox{.}(2023)]%
        {hou2023large}
\bibfield{author}{\bibinfo{person}{Xinyi Hou}, \bibinfo{person}{Yanjie Zhao},
  \bibinfo{person}{Yue Liu}, \bibinfo{person}{Zhou Yang},
  \bibinfo{person}{Kailong Wang}, \bibinfo{person}{Li Li},
  \bibinfo{person}{Xiapu Luo}, \bibinfo{person}{David Lo},
  \bibinfo{person}{John Grundy}, {and} \bibinfo{person}{Haoyu Wang}.}
  \bibinfo{year}{2023}\natexlab{}.
\newblock \bibinfo{title}{Large Language Models for Software Engineering: A
  Systematic Literature Review}.
\newblock
\newblock
\showeprint[arxiv]{2308.10620}~[cs.SE]


\bibitem[Imai(2022)]%
        {Imai-2022-copilot:ICSE22}
\bibfield{author}{\bibinfo{person}{Saki Imai}.}
  \bibinfo{year}{2022}\natexlab{}.
\newblock \showarticletitle{Is GitHub Copilot a Substitute for Human
  Pair-Programming? An Empirical Study}. In
  \bibinfo{booktitle}{\emph{Proceedings of the ACM/IEEE 44th International
  Conference on Software Engineering: Companion Proceedings}} (Pittsburgh,
  Pennsylvania) \emph{(\bibinfo{series}{ICSE '22})}.
  \bibinfo{publisher}{Association for Computing Machinery},
  \bibinfo{address}{New York, NY, USA}, \bibinfo{pages}{319–321}.
\newblock
\showISBNx{9781450392235}
\urldef\tempurl%
\url{https://doi.org/10.1145/3510454.3522684}
\showDOI{\tempurl}


\bibitem[Kaufeld et~al\mbox{.}(2022)]%
        {ar-induce-sickness:2022}
\bibfield{author}{\bibinfo{person}{Mara Kaufeld}, \bibinfo{person}{Martin
  Mundt}, \bibinfo{person}{Sarah Forst}, {and} \bibinfo{person}{Heiko Hecht}.}
  \bibinfo{year}{2022}\natexlab{}.
\newblock \showarticletitle{Optical see-through augmented reality can induce
  severe motion sickness}.
\newblock \bibinfo{journal}{\emph{Displays}}  \bibinfo{volume}{74}
  (\bibinfo{year}{2022}), \bibinfo{pages}{102283}.
\newblock
\urldef\tempurl%
\url{https://doi.org/10.1016/j.displa.2022.102283}
\showDOI{\tempurl}


\bibitem[Krause-Glau et~al\mbox{.}(2022)]%
        {krause2022collaborative}
\bibfield{author}{\bibinfo{person}{Alexander Krause-Glau},
  \bibinfo{person}{Malte Hansen}, {and} \bibinfo{person}{Wilhelm Hasselbring}.}
  \bibinfo{year}{2022}\natexlab{}.
\newblock \showarticletitle{Collaborative program comprehension via software
  visualization in extended reality}.
\newblock \bibinfo{journal}{\emph{Information and Software Technology}}
  \bibinfo{volume}{151} (\bibinfo{year}{2022}), \bibinfo{pages}{107007}.
\newblock


\bibitem[Krau{\ss} et~al\mbox{.}(2021)]%
        {krauss2021current}
\bibfield{author}{\bibinfo{person}{Veronika Krau{\ss}},
  \bibinfo{person}{Alexander Boden}, \bibinfo{person}{Leif Oppermann}, {and}
  \bibinfo{person}{Ren{\'e} Reiners}.} \bibinfo{year}{2021}\natexlab{}.
\newblock \showarticletitle{Current practices, challenges, and design
  implications for collaborative ar/vr application development}. In
  \bibinfo{booktitle}{\emph{Proceedings of the 2021 CHI Conference on Human
  Factors in Computing Systems}}. \bibinfo{pages}{1--15}.
\newblock


\bibitem[Lorey et~al\mbox{.}(2022)]%
        {lorey-theories22}
\bibfield{author}{\bibinfo{person}{Tobias Lorey}, \bibinfo{person}{Paul Ralph},
  {and} \bibinfo{person}{Michael Felderer}.} \bibinfo{year}{2022}\natexlab{}.
\newblock \showarticletitle{Social Science Theories in Software Engineering
  Research}. In \bibinfo{booktitle}{\emph{Proceedings of the 44th International
  Conference on Software Engineering}} (Pittsburgh, Pennsylvania)
  \emph{(\bibinfo{series}{ICSE '22})}. \bibinfo{publisher}{Association for
  Computing Machinery}, \bibinfo{address}{New York, NY, USA},
  \bibinfo{pages}{1994–2005}.
\newblock
\showISBNx{9781450392211}
\urldef\tempurl%
\url{https://doi.org/10.1145/3510003.3510076}
\showDOI{\tempurl}


\bibitem[Lotufo et~al\mbox{.}(2012)]%
        {Lotufo:etAl:BugTracking}
\bibfield{author}{\bibinfo{person}{Rafael Lotufo},
  \bibinfo{person}{Leonardo~Teixeira Passos}, {and} \bibinfo{person}{Krzysztof
  Czarnecki}.} \bibinfo{year}{2012}\natexlab{}.
\newblock \showarticletitle{Towards improving bug tracking systems with game
  mechanisms}. In \bibinfo{booktitle}{\emph{9th {IEEE} Working Conference of
  Mining Software Repositories, {MSR} 2012, June 2-3, 2012, Zurich,
  Switzerland}}, \bibfield{editor}{\bibinfo{person}{Michele Lanza},
  \bibinfo{person}{Massimiliano~Di Penta}, {and} \bibinfo{person}{Tao Xie}}
  (Eds.). \bibinfo{publisher}{{IEEE} Computer Society}, \bibinfo{pages}{2--11}.
\newblock
\urldef\tempurl%
\url{https://doi.org/10.1109/MSR.2012.6224293}
\showDOI{\tempurl}


\bibitem[Lukasik(2011)]%
        {ARPANET}
\bibfield{author}{\bibinfo{person}{Stephen Lukasik}.}
  \bibinfo{year}{2011}\natexlab{}.
\newblock \showarticletitle{Why the Arpanet Was Built}.
\newblock \bibinfo{journal}{\emph{IEEE Annals of the History of Computing}}
  \bibinfo{volume}{33}, \bibinfo{number}{3} (\bibinfo{year}{2011}),
  \bibinfo{pages}{4--21}.
\newblock
\urldef\tempurl%
\url{https://doi.org/10.1109/MAHC.2010.11}
\showDOI{\tempurl}


\bibitem[Mamykina et~al\mbox{.}(2011)]%
        {Mamykina:etAl:designLessons}
\bibfield{author}{\bibinfo{person}{Lena Mamykina}, \bibinfo{person}{Bella
  Manoim}, \bibinfo{person}{Manas Mittal}, \bibinfo{person}{George Hripcsak},
  {and} \bibinfo{person}{Bj\"{o}rn Hartmann}.} \bibinfo{year}{2011}\natexlab{}.
\newblock \showarticletitle{Design Lessons from the Fastest Q\&a Site in the
  West}. In \bibinfo{booktitle}{\emph{Proceedings of the SIGCHI Conference on
  Human Factors in Computing Systems}} (Vancouver, BC, Canada)
  \emph{(\bibinfo{series}{CHI '11})}. \bibinfo{publisher}{Association for
  Computing Machinery}, \bibinfo{address}{New York, NY, USA},
  \bibinfo{pages}{2857–2866}.
\newblock
\showISBNx{9781450302289}
\urldef\tempurl%
\url{https://doi.org/10.1145/1978942.1979366}
\showDOI{\tempurl}


\bibitem[McGrath(1995)]%
        {mcgrath1995methodology}
\bibfield{author}{\bibinfo{person}{Joseph~E McGrath}.}
  \bibinfo{year}{1995}\natexlab{}.
\newblock \showarticletitle{Methodology matters: Doing research in the
  behavioral and social sciences}.
\newblock In \bibinfo{booktitle}{\emph{Readings in Human--Computer
  Interaction}}. \bibinfo{publisher}{Elsevier}, \bibinfo{pages}{152--169}.
\newblock


\bibitem[McLuhan(1977)]%
        {mcluhan77}
\bibfield{author}{\bibinfo{person}{Marshall McLuhan}.}
  \bibinfo{year}{1977}\natexlab{}.
\newblock \showarticletitle{Laws of the Media}.
\newblock \bibinfo{journal}{\emph{ETC: A Review of General Semantics}}
  (\bibinfo{year}{1977}), \bibinfo{pages}{173--179}.
\newblock


\bibitem[McLuhan(2017)]%
        {mcluhan2017medium}
\bibfield{author}{\bibinfo{person}{Marshall McLuhan}.}
  \bibinfo{year}{2017}\natexlab{}.
\newblock \showarticletitle{The medium is the message}.
\newblock In \bibinfo{booktitle}{\emph{Communication theory}}.
  \bibinfo{publisher}{Routledge}, \bibinfo{pages}{390--402}.
\newblock


\bibitem[Meldrum et~al\mbox{.}(2020b)]%
        {MELDRUM2020102516}
\bibfield{author}{\bibinfo{person}{Sarah Meldrum}, \bibinfo{person}{Sherlock~A.
  Licorish}, \bibinfo{person}{Caitlin~A. Owen}, {and} \bibinfo{person}{Bastin
  Tony~Roy Savarimuthu}.} \bibinfo{year}{2020}\natexlab{b}.
\newblock \showarticletitle{Understanding stack overflow code quality: A
  recommendation of caution}.
\newblock \bibinfo{journal}{\emph{Science of Computer Programming}}
  \bibinfo{volume}{199} (\bibinfo{year}{2020}), \bibinfo{pages}{102516}.
\newblock
\showISSN{0167-6423}
\urldef\tempurl%
\url{https://doi.org/10.1016/j.scico.2020.102516}
\showDOI{\tempurl}


\bibitem[Meldrum et~al\mbox{.}(2020a)]%
        {meldrum2020}
\bibfield{author}{\bibinfo{person}{Sarah Meldrum}, \bibinfo{person}{Sherlock~A.
  Licorish}, {and} \bibinfo{person}{Bastin Tony~Roy Savarimuthu}.}
  \bibinfo{year}{2020}\natexlab{a}.
\newblock \showarticletitle{Exploring Research Interest in Stack Overflow - {A}
  Systematic Mapping Study and Quality Evaluation}.
\newblock \bibinfo{journal}{\emph{CoRR}}  \bibinfo{volume}{abs/2010.12282}
  (\bibinfo{year}{2020}).
\newblock
\showeprint[arXiv]{2010.12282}
\urldef\tempurl%
\url{https://arxiv.org/abs/2010.12282}
\showURL{%
\tempurl}


\bibitem[Merchant et~al\mbox{.}(2019)]%
        {Merchant:popularity:2019}
\bibfield{author}{\bibinfo{person}{Arpit Merchant}, \bibinfo{person}{Daksh
  Shah}, \bibinfo{person}{Gurpreet~Singh Bhatia}, \bibinfo{person}{Anurag
  Ghosh}, {and} \bibinfo{person}{Ponnurangam Kumaraguru}.}
  \bibinfo{year}{2019}\natexlab{}.
\newblock \showarticletitle{Signals Matter: Understanding Popularity and Impact
  of Users on Stack Overflow}. In \bibinfo{booktitle}{\emph{The World Wide Web
  Conference}} (San Francisco, CA, USA) \emph{(\bibinfo{series}{WWW '19})}.
  \bibinfo{publisher}{Association for Computing Machinery},
  \bibinfo{address}{New York, NY, USA}, \bibinfo{pages}{3086–3092}.
\newblock
\showISBNx{9781450366748}
\urldef\tempurl%
\url{https://doi.org/10.1145/3308558.3313583}
\showDOI{\tempurl}


\bibitem[Mokyr(1992)]%
        {mokyr1992lever}
\bibfield{author}{\bibinfo{person}{Joel Mokyr}.}
  \bibinfo{year}{1992}\natexlab{}.
\newblock \bibinfo{booktitle}{\emph{The lever of riches: Technological
  creativity and economic progress}}.
\newblock \bibinfo{publisher}{Oxford University Press}.
\newblock


\bibitem[Moutidis and Williams(2021)]%
        {moutidis2021community}
\bibfield{author}{\bibinfo{person}{Iraklis Moutidis} {and}
  \bibinfo{person}{Hywel~TP Williams}.} \bibinfo{year}{2021}\natexlab{}.
\newblock \showarticletitle{Community evolution on stack overflow}.
\newblock \bibinfo{journal}{\emph{Plos one}} \bibinfo{volume}{16},
  \bibinfo{number}{6} (\bibinfo{year}{2021}), \bibinfo{pages}{e0253010}.
\newblock


\bibitem[Murgia et~al\mbox{.}(2016)]%
        {Murgia:bots:SO}
\bibfield{author}{\bibinfo{person}{Alessandro Murgia}, \bibinfo{person}{Daan
  Janssens}, \bibinfo{person}{Serge Demeyer}, {and} \bibinfo{person}{Bogdan
  Vasilescu}.} \bibinfo{year}{2016}\natexlab{}.
\newblock \showarticletitle{Among the Machines: Human-Bot Interaction on Social
  Q\&A Websites}. In \bibinfo{booktitle}{\emph{Proceedings of the 2016 CHI
  Conference Extended Abstracts on Human Factors in Computing Systems}} (San
  Jose, California, USA) \emph{(\bibinfo{series}{CHI EA '16})}.
  \bibinfo{publisher}{Association for Computing Machinery},
  \bibinfo{address}{New York, NY, USA}, \bibinfo{pages}{1272–1279}.
\newblock
\showISBNx{9781450340823}
\urldef\tempurl%
\url{https://doi.org/10.1145/2851581.2892311}
\showDOI{\tempurl}


\bibitem[Murphy et~al\mbox{.}(2006)]%
        {murphy2006java}
\bibfield{author}{\bibinfo{person}{Gail Murphy}, \bibinfo{person}{Mik Kersten},
  {and} \bibinfo{person}{Leah Findlater}.} \bibinfo{year}{2006}\natexlab{}.
\newblock \showarticletitle{How are java software developers using the eclipse
  ide?}
\newblock \bibinfo{journal}{\emph{IEEE Software}} \bibinfo{volume}{23},
  \bibinfo{number}{4} (\bibinfo{year}{2006}), \bibinfo{pages}{76--83}.
\newblock


\bibitem[Novielli et~al\mbox{.}(2014)]%
        {novielli2014towards}
\bibfield{author}{\bibinfo{person}{Nicole Novielli}, \bibinfo{person}{Fabio
  Calefato}, {and} \bibinfo{person}{Filippo Lanubile}.}
  \bibinfo{year}{2014}\natexlab{}.
\newblock \showarticletitle{Towards discovering the role of emotions in stack
  overflow}. In \bibinfo{booktitle}{\emph{Proceedings of the 6th international
  workshop on social software engineering}}. \bibinfo{pages}{33--36}.
\newblock


\bibitem[OpenAI(2023)]%
        {openai2023gpt4}
\bibfield{author}{\bibinfo{person}{OpenAI}.} \bibinfo{year}{2023}\natexlab{}.
\newblock \bibinfo{title}{GPT-4 Technical Report}.
\newblock
\newblock
\showeprint[arxiv]{2303.08774}~[cs.CL]


\bibitem[Parnin and Orso(2011)]%
        {parnin2011automated}
\bibfield{author}{\bibinfo{person}{Chris Parnin} {and}
  \bibinfo{person}{Alessandro Orso}.} \bibinfo{year}{2011}\natexlab{}.
\newblock \showarticletitle{Are automated debugging techniques actually helping
  programmers?}. In \bibinfo{booktitle}{\emph{Proceedings of the International
  Symposium on Software Testing and Analysis}}. ACM.
\newblock


\bibitem[Parnin et~al\mbox{.}(2012)]%
        {parnin2012crowd}
\bibfield{author}{\bibinfo{person}{Chris Parnin}, \bibinfo{person}{Christoph
  Treude}, \bibinfo{person}{Lars Grammel}, {and} \bibinfo{person}{Margaret-Anne
  Storey}.} \bibinfo{year}{2012}\natexlab{}.
\newblock \showarticletitle{Crowd documentation: Exploring the coverage and the
  dynamics of API discussions on Stack Overflow}.
\newblock \bibinfo{journal}{\emph{Georgia Institute of Technology, Tech. Rep}}
  \bibinfo{volume}{11} (\bibinfo{year}{2012}).
\newblock


\bibitem[Peng et~al\mbox{.}(2023)]%
        {Peng:copilot:controlledStudy}
\bibfield{author}{\bibinfo{person}{Sida Peng}, \bibinfo{person}{Eirini
  Kalliamvakou}, \bibinfo{person}{Peter Cihon}, {and} \bibinfo{person}{Mert
  Demirere}.} \bibinfo{year}{2023}\natexlab{}.
\newblock \showarticletitle{The Impact of AI on Developer Productivity:
  Evidence from GitHub}.
\newblock \bibinfo{journal}{\emph{arXiv preprint arXiv:2302.06590}}
  (\bibinfo{year}{2023}).
\newblock
\urldef\tempurl%
\url{http://arxiv.org/abs/2302.06590}
\showURL{%
\tempurl}


\bibitem[Perry(2016)]%
        {Perry:2016}
\bibfield{author}{\bibinfo{person}{Tekla~S. Perry}.}
  \bibinfo{year}{2016}\natexlab{}.
\newblock \showarticletitle{Virtual reality goes social}.
\newblock \bibinfo{journal}{\emph{IEEE Spectrum}} \bibinfo{volume}{53},
  \bibinfo{number}{1} (\bibinfo{year}{2016}), \bibinfo{pages}{56--57}.
\newblock
\urldef\tempurl%
\url{https://doi.org/10.1109/MSPEC.2016.7367470}
\showDOI{\tempurl}


\bibitem[P{\"o}ial(2021)]%
        {poial2021}
\bibfield{author}{\bibinfo{person}{Jaanus P{\"o}ial}.}
  \bibinfo{year}{2021}\natexlab{}.
\newblock \showarticletitle{Challenges of Teaching Programming in StackOverflow
  Era}. In \bibinfo{booktitle}{\emph{Educating Engineers for Future Industrial
  Revolutions}}, \bibfield{editor}{\bibinfo{person}{Michael~E. Auer} {and}
  \bibinfo{person}{Tiia R{\"u}{\"u}tmann}} (Eds.). \bibinfo{publisher}{Springer
  International Publishing}, \bibinfo{address}{Cham},
  \bibinfo{pages}{703--710}.
\newblock
\showISBNx{978-3-030-68198-2}


\bibitem[Ponzanelli et~al\mbox{.}(2013)]%
        {ponzanelli2013seahawk}
\bibfield{author}{\bibinfo{person}{Luca Ponzanelli}, \bibinfo{person}{Alberto
  Bacchelli}, {and} \bibinfo{person}{Michele Lanza}.}
  \bibinfo{year}{2013}\natexlab{}.
\newblock \showarticletitle{Seahawk: Stack overflow in the ide}. In
  \bibinfo{booktitle}{\emph{2013 35th International Conference on Software
  Engineering (ICSE)}}. IEEE, \bibinfo{pages}{1295--1298}.
\newblock


\bibitem[Ragkhitwetsagul et~al\mbox{.}(2021)]%
        {Ragkhitwetsagul:et:AL:toxicCode:SO}
\bibfield{author}{\bibinfo{person}{Chaiyong Ragkhitwetsagul},
  \bibinfo{person}{Jens Krinke}, \bibinfo{person}{Matheus Paixao},
  \bibinfo{person}{Giuseppe Bianco}, {and} \bibinfo{person}{Rocco Oliveto}.}
  \bibinfo{year}{2021}\natexlab{}.
\newblock \showarticletitle{Toxic Code Snippets on Stack Overflow}.
\newblock \bibinfo{journal}{\emph{IEEE Transactions on Software Engineering}}
  \bibinfo{volume}{47}, \bibinfo{number}{3} (\bibinfo{year}{2021}),
  \bibinfo{pages}{560--581}.
\newblock
\urldef\tempurl%
\url{https://doi.org/10.1109/TSE.2019.2900307}
\showDOI{\tempurl}


\bibitem[Ralph et~al\mbox{.}(2020b)]%
        {ralph2020acm}
\bibfield{author}{\bibinfo{person}{P. Ralph} {et~al\mbox{.}}}
  \bibinfo{year}{2020}\natexlab{b}.
\newblock \showarticletitle{ACM SIGSOFT Empirical Standards}.
\newblock \bibinfo{journal}{\emph{arXiv preprint arXiv:2010.03525}}
  (\bibinfo{year}{2020}).
\newblock


\bibitem[Ralph et~al\mbox{.}(2020a)]%
        {ralph2020empirical}
\bibfield{author}{\bibinfo{person}{Paul Ralph}, \bibinfo{person}{Nauman~bin
  Ali}, \bibinfo{person}{Sebastian Baltes}, \bibinfo{person}{Domenico
  Bianculli}, \bibinfo{person}{Jessica Diaz}, \bibinfo{person}{Yvonne
  Dittrich}, \bibinfo{person}{Neil Ernst}, \bibinfo{person}{Michael Felderer},
  \bibinfo{person}{Robert Feldt}, \bibinfo{person}{Antonio Filieri},
  {et~al\mbox{.}}} \bibinfo{year}{2020}\natexlab{a}.
\newblock \showarticletitle{Empirical standards for software engineering
  research}.
\newblock \bibinfo{journal}{\emph{arXiv preprint arXiv:2010.03525}}
  (\bibinfo{year}{2020}).
\newblock


\bibitem[Ross et~al\mbox{.}(2023)]%
        {ross2023programmer}
\bibfield{author}{\bibinfo{person}{Steven~I Ross}, \bibinfo{person}{Fernando
  Martinez}, \bibinfo{person}{Stephanie Houde}, \bibinfo{person}{Michael
  Muller}, {and} \bibinfo{person}{Justin~D Weisz}.}
  \bibinfo{year}{2023}\natexlab{}.
\newblock \showarticletitle{The programmer’s assistant: Conversational
  interaction with a large language model for software development}. In
  \bibinfo{booktitle}{\emph{Proceedings of the 28th International Conference on
  Intelligent User Interfaces}}. \bibinfo{pages}{491--514}.
\newblock


\bibitem[Silva et~al\mbox{.}(2021)]%
        {SILVA2021111063}
\bibfield{author}{\bibinfo{person}{Rodrigo~F. Silva},
  \bibinfo{person}{Mohammad~Masudur Rahman}, \bibinfo{person}{Carlos~Eduardo
  Dantas}, \bibinfo{person}{Chanchal Roy}, \bibinfo{person}{Foutse Khomh},
  {and} \bibinfo{person}{Marcelo~A. Maia}.} \bibinfo{year}{2021}\natexlab{}.
\newblock \showarticletitle{Improved retrieval of programming solutions with
  code examples using a multi-featured score}.
\newblock \bibinfo{journal}{\emph{Journal of Systems and Software}}
  \bibinfo{volume}{181} (\bibinfo{year}{2021}), \bibinfo{pages}{111063}.
\newblock
\showISSN{0164-1212}
\urldef\tempurl%
\url{https://doi.org/10.1016/j.jss.2021.111063}
\showDOI{\tempurl}


\bibitem[Sj{\o}berg et~al\mbox{.}(2008)]%
        {Sjoberg-theories2008}
\bibfield{author}{\bibinfo{person}{Dag I.~K. Sj{\o}berg}, \bibinfo{person}{Tore
  Dyb{\aa}}, \bibinfo{person}{Bente C.~D. Anda}, {and} \bibinfo{person}{Jo~E.
  Hannay}.} \bibinfo{year}{2008}\natexlab{}.
\newblock \bibinfo{booktitle}{\emph{Building Theories in Software
  Engineering}}.
\newblock \bibinfo{publisher}{Springer London}, \bibinfo{address}{London},
  \bibinfo{pages}{312--336}.
\newblock
\showISBNx{978-1-84800-044-5}
\urldef\tempurl%
\url{https://doi.org/10.1007/978-1-84800-044-5_12}
\showDOI{\tempurl}


\bibitem[Squire(2015a)]%
        {squire2015should}
\bibfield{author}{\bibinfo{person}{Megan Squire}.}
  \bibinfo{year}{2015}\natexlab{a}.
\newblock \showarticletitle{" Should We Move to Stack Overflow?" Measuring the
  Utility of Social Media for Developer Support}. In
  \bibinfo{booktitle}{\emph{2015 IEEE/ACM 37th IEEE International Conference on
  Software Engineering}}, Vol.~\bibinfo{volume}{2}. IEEE,
  \bibinfo{pages}{219--228}.
\newblock


\bibitem[Squire(2015b)]%
        {Squire:moving_to_SO}
\bibfield{author}{\bibinfo{person}{Megan Squire}.}
  \bibinfo{year}{2015}\natexlab{b}.
\newblock \showarticletitle{Should We Move to Stack Overflow? Measuring the
  Utility of Social Media for Developer Support}. In
  \bibinfo{booktitle}{\emph{Proceedings of the 37th International Conference on
  Software Engineering - Volume 2}} (Florence, Italy)
  \emph{(\bibinfo{series}{ICSE '15})}. \bibinfo{publisher}{IEEE Press},
  \bibinfo{pages}{219–228}.
\newblock


\bibitem[Squire and Funkhouser(2014)]%
        {squire2014}
\bibfield{author}{\bibinfo{person}{Megan Squire} {and}
  \bibinfo{person}{Christian Funkhouser}.} \bibinfo{year}{2014}\natexlab{}.
\newblock \showarticletitle{"A Bit of Code": How the Stack Overflow Community
  Creates Quality Postings}. In \bibinfo{booktitle}{\emph{2014 47th Hawaii
  International Conference on System Sciences}}. \bibinfo{pages}{1425--1434}.
\newblock
\urldef\tempurl%
\url{https://doi.org/10.1109/HICSS.2014.185}
\showDOI{\tempurl}


\bibitem[Stol and Fitzgerald(2013)]%
        {stol-theories13}
\bibfield{author}{\bibinfo{person}{Klaas-Jan Stol} {and} \bibinfo{person}{Brian
  Fitzgerald}.} \bibinfo{year}{2013}\natexlab{}.
\newblock \showarticletitle{Uncovering theories in software engineering}. In
  \bibinfo{booktitle}{\emph{2013 2nd SEMAT Workshop on a General Theory of
  Software Engineering (GTSE)}}. \bibinfo{pages}{5--14}.
\newblock
\urldef\tempurl%
\url{https://doi.org/10.1109/GTSE.2013.6613863}
\showDOI{\tempurl}


\bibitem[Storey et~al\mbox{.}(2020)]%
        {storey2020software}
\bibfield{author}{\bibinfo{person}{Margaret-Anne Storey},
  \bibinfo{person}{Neil~A Ernst}, \bibinfo{person}{Courtney Williams}, {and}
  \bibinfo{person}{Eirini Kalliamvakou}.} \bibinfo{year}{2020}\natexlab{}.
\newblock \showarticletitle{The who, what, how of software engineering
  research: a socio-technical framework}.
\newblock \bibinfo{journal}{\emph{Empirical Software Engineering}}
  \bibinfo{volume}{25} (\bibinfo{year}{2020}), \bibinfo{pages}{4097--4129}.
\newblock


\bibitem[Subramanian and Holmes(2013)]%
        {Subramanian2013Making}
\bibfield{author}{\bibinfo{person}{Siddharth Subramanian} {and}
  \bibinfo{person}{Reid Holmes}.} \bibinfo{year}{2013}\natexlab{}.
\newblock \showarticletitle{Making sense of online code snippets}. In
  \bibinfo{booktitle}{\emph{Proceedings of the 10th International Working
  Conference on Mining Software Repositories}}. \bibinfo{publisher}{IEEE},
  \bibinfo{pages}{85--88}.
\newblock


\bibitem[Tang and Nadi(2021)]%
        {Tang:Nadi:2021}
\bibfield{author}{\bibinfo{person}{Henry Tang} {and} \bibinfo{person}{Sarah
  Nadi}.} \bibinfo{year}{2021}\natexlab{}.
\newblock \showarticletitle{On using Stack Overflow comment-edit pairs to
  recommend code maintenance changes}.
\newblock \bibinfo{journal}{\emph{Empirical Softw. Engg.}}
  \bibinfo{volume}{26}, \bibinfo{number}{4} (\bibinfo{date}{jul}
  \bibinfo{year}{2021}), \bibinfo{numpages}{35}~pages.
\newblock
\showISSN{1382-3256}
\urldef\tempurl%
\url{https://doi.org/10.1007/s10664-021-09954-8}
\showDOI{\tempurl}


\bibitem[Thomas(2007)]%
        {mcluhan_operationalizing_2007}
\bibfield{author}{\bibinfo{person}{John Thomas}.}
  \bibinfo{year}{2007}\natexlab{}.
\newblock \showarticletitle{Operationalizing {McLuhan}'s tetrad to focus on
  innovation effects}.
\newblock  (\bibinfo{date}{Jan.} \bibinfo{year}{2007}).
\newblock
\urldef\tempurl%
\url{https://dr.lib.iastate.edu/handle/20.500.12876/68461}
\showURL{%
\tempurl}


\bibitem[Touvron et~al\mbox{.}(2023)]%
        {touvron2023llama}
\bibfield{author}{\bibinfo{person}{Hugo Touvron}, \bibinfo{person}{Thibaut
  Lavril}, \bibinfo{person}{Gautier Izacard}, \bibinfo{person}{Xavier
  Martinet}, \bibinfo{person}{Marie-Anne Lachaux}, \bibinfo{person}{Timothée
  Lacroix}, \bibinfo{person}{Baptiste Rozière}, \bibinfo{person}{Naman Goyal},
  \bibinfo{person}{Eric Hambro}, \bibinfo{person}{Faisal Azhar},
  \bibinfo{person}{Aurelien Rodriguez}, \bibinfo{person}{Armand Joulin},
  \bibinfo{person}{Edouard Grave}, {and} \bibinfo{person}{Guillaume Lample}.}
  \bibinfo{year}{2023}\natexlab{}.
\newblock \bibinfo{title}{LLaMA: Open and Efficient Foundation Language
  Models}.
\newblock
\newblock
\showeprint[arxiv]{2302.13971}~[cs.CL]


\bibitem[Treude et~al\mbox{.}(2011)]%
        {Treude:etAl:AskingQuestions}
\bibfield{author}{\bibinfo{person}{Christoph Treude}, \bibinfo{person}{Ohad
  Barzilay}, {and} \bibinfo{person}{Margaret{-}Anne~D. Storey}.}
  \bibinfo{year}{2011}\natexlab{}.
\newblock \showarticletitle{How do programmers ask and answer questions on the
  web?}. In \bibinfo{booktitle}{\emph{Proceedings of the 33rd International
  Conference on Software Engineering, {ICSE} 2011, Waikiki, Honolulu , HI, USA,
  May 21-28, 2011}}, \bibfield{editor}{\bibinfo{person}{Richard~N. Taylor},
  \bibinfo{person}{Harald~C. Gall}, {and} \bibinfo{person}{Nenad Medvidovic}}
  (Eds.). \bibinfo{publisher}{{ACM}}, \bibinfo{pages}{804--807}.
\newblock
\urldef\tempurl%
\url{https://doi.org/10.1145/1985793.1985907}
\showDOI{\tempurl}


\bibitem[Treude and Robillard(2016)]%
        {treude2016augmenting}
\bibfield{author}{\bibinfo{person}{Christoph Treude} {and}
  \bibinfo{person}{Martin~P Robillard}.} \bibinfo{year}{2016}\natexlab{}.
\newblock \showarticletitle{Augmenting API documentation with insights from
  stack overflow}. In \bibinfo{booktitle}{\emph{Proceedings of the 38th
  International Conference on Software Engineering}}.
  \bibinfo{pages}{392--403}.
\newblock


\bibitem[Treude and Robillard(2017)]%
        {Treude:Robillard:2017:understsandingSO}
\bibfield{author}{\bibinfo{person}{Christoph Treude} {and}
  \bibinfo{person}{Martin~P. Robillard}.} \bibinfo{year}{2017}\natexlab{}.
\newblock \showarticletitle{Understanding Stack Overflow Code Fragments}. In
  \bibinfo{booktitle}{\emph{2017 IEEE International Conference on Software
  Maintenance and Evolution (ICSME)}}. \bibinfo{pages}{509--513}.
\newblock
\urldef\tempurl%
\url{https://doi.org/10.1109/ICSME.2017.24}
\showDOI{\tempurl}


\bibitem[Vasilescu et~al\mbox{.}(2013)]%
        {Vasilescu2013Associations}
\bibfield{author}{\bibinfo{person}{Bogdan Vasilescu}, \bibinfo{person}{Vladimir
  Filkov}, {and} \bibinfo{person}{Alexander Serebrenik}.}
  \bibinfo{year}{2013}\natexlab{}.
\newblock \showarticletitle{{StackOverflow} and {GitHub}: {A}ssociations
  between software development and crowdsourced knowledge}. In
  \bibinfo{booktitle}{\emph{Proceedings of the 2013 ASE/IEEE International
  Conference on Social Computing}} (Washington D.C., USA).
  \bibinfo{publisher}{IEEE}, \bibinfo{pages}{188--195}.
\newblock


\bibitem[Wagner and Ruhe(2018)]%
        {wagner2018systematic}
\bibfield{author}{\bibinfo{person}{Stefan Wagner} {and}
  \bibinfo{person}{G{\"u}nter Ruhe}.} \bibinfo{year}{2018}\natexlab{}.
\newblock \showarticletitle{A systematic review of productivity factors in
  software development}.
\newblock \bibinfo{journal}{\emph{Arxiv}} (\bibinfo{year}{2018}).
\newblock
\urldef\tempurl%
\url{https://arxiv.org/abs/1801.06475}
\showURL{%
\tempurl}


\bibitem[Wohlin et~al\mbox{.}(2012)]%
        {wohlin2012experimentation}
\bibfield{author}{\bibinfo{person}{C. Wohlin}, \bibinfo{person}{P. Runeson},
  \bibinfo{person}{M. H{\"o}st}, \bibinfo{person}{M. Ohlsson},
  \bibinfo{person}{B. Regnell}, {and} \bibinfo{person}{A. Wessl{\'e}n}.}
  \bibinfo{year}{2012}\natexlab{}.
\newblock \bibinfo{booktitle}{\emph{Experimentation in Software Engineering}}.
\newblock \bibinfo{publisher}{Springer Science \& Business Media}.
\newblock


\bibitem[Wu et~al\mbox{.}(2019)]%
        {Yuhao:etAl:2019:understanding:reuse}
\bibfield{author}{\bibinfo{person}{Yuhao Wu}, \bibinfo{person}{Shaowei Wang},
  \bibinfo{person}{Cor-Paul Bezemer}, {and} \bibinfo{person}{Katsuro Inoue}.}
  \bibinfo{year}{2019}\natexlab{}.
\newblock \showarticletitle{How Do Developers Utilize Source Code from Stack
  Overflow?}
\newblock \bibinfo{journal}{\emph{Empirical Softw. Engg.}}
  \bibinfo{volume}{24}, \bibinfo{number}{2} (\bibinfo{date}{apr}
  \bibinfo{year}{2019}), \bibinfo{pages}{637–673}.
\newblock
\showISSN{1382-3256}
\urldef\tempurl%
\url{https://doi.org/10.1007/s10664-018-9634-5}
\showDOI{\tempurl}


\bibitem[Yilmaz et~al\mbox{.}(2023)]%
        {yilmaz2023examining}
\bibfield{author}{\bibinfo{person}{Murat Yilmaz}, \bibinfo{person}{Emer
  O'farrell}, {and} \bibinfo{person}{Paul Clarke}.}
  \bibinfo{year}{2023}\natexlab{}.
\newblock \showarticletitle{Examining the training and education potential of
  the metaverse: Results from an empirical study of next generation SAFe
  training}.
\newblock \bibinfo{journal}{\emph{Journal of Software: Evolution and Process}}
  (\bibinfo{year}{2023}), \bibinfo{pages}{e2531}.
\newblock


\bibitem[Zhang et~al\mbox{.}(2018)]%
        {Zhang:etAL:reliableSO:2018}
\bibfield{author}{\bibinfo{person}{Tianyi Zhang}, \bibinfo{person}{Ganesha
  Upadhyaya}, \bibinfo{person}{Anastasia Reinhardt}, \bibinfo{person}{Hridesh
  Rajan}, {and} \bibinfo{person}{Miryung Kim}.}
  \bibinfo{year}{2018}\natexlab{}.
\newblock \showarticletitle{Are Code Examples on an Online Q\&A Forum Reliable?
  A Study of API Misuse on Stack Overflow}. In
  \bibinfo{booktitle}{\emph{Proceedings of the 40th International Conference on
  Software Engineering}} (Gothenburg, Sweden) \emph{(\bibinfo{series}{ICSE
  '18})}. \bibinfo{publisher}{Association for Computing Machinery},
  \bibinfo{address}{New York, NY, USA}, \bibinfo{pages}{886–896}.
\newblock
\showISBNx{9781450356381}
\urldef\tempurl%
\url{https://doi.org/10.1145/3180155.3180260}
\showDOI{\tempurl}


\end{thebibliography}

\appendix


\section{The MyResearchPlaybookGPT: Enhancing Research on Disruptive Technologies}
\label{ResearchPlaybookGPT}

Software engineering is continuously being reshaped by disruptive technologies. These innovations necessitate a multifaceted approach to research, one that comprehends both the technological advancements and their socio-technical implications. In this regard, we introduce ``MyResearchPlaybook,'' a specialized GPT model, as a significant adjunct to our research playbook and to help in particular in the brainstorming activity the playbook invokes. In this section, we describe a playbook GPT we have designed to assist in using the playbook. In Section~\ref{sec:examples_ARVR} we show how we applied this GPT to assist (not replace!) our activity of applying the playbook to the role of AR/VR in SE. 

\subsection{GPTs: Tailored AI for Focused Inquiry}

GPTs, a subset of the broader Generative Pre-trained Transformer models by OpenAI, are customizable AI tools designed to address specific challenges and tasks. These models extend the capabilities of standard GPTs by focusing on particular domains or applications, offering more nuanced and relevant insights. An example of this is MyResearchPlaybook\footnote{MyResearchPlaybook is accessible at this link in OpenAI's library: \url{https://chat.openai.com/g/g-ckghjdz0n-myresearchplaybook}.}, a GPT model configured to assist in exploring the interplay between disruptive technologies and their human-centric impacts. To make the instructions transparent and reusable with other LLMs, we published them in a GitHub repo\footnote{Link to the GitHub repo with the instructions: \url{https://github.com/danielrusso-org/MyResearchPlaybook}.}.

\subsection{MyResearchPlaybook: Bridging McLuhan's Theory and Software Engineering Research}

Drawing upon the theoretical framework of McLuhan's Triadic Dimensions---enhances, obsoletes, retrieves, and reverses---MyResearchPlaybook aids in constructing a structured matrix for analysis. This matrix serves as a scaffold for developing research questions that are pivotal in understanding the multifaceted effects of disruptive technologies such as AI, AR, VR, and their iterations in software engineering.

\subsection{Capabilities and Contributions}

\begin{itemize}
    \item \textbf{Analytical Matrix Construction}: MyResearchPlaybook generates matrices aligning McLuhan's Dimensions with current research phenomena in software engineering, ensuring a comprehensive exploration of potential impacts.
    \item \textbf{Formulation of Research Questions}: It proposes specific research questions at the intersection of each dimension and phenomenon, guiding researchers toward targeted areas of exploration.
    \item \textbf{Strategic Research Planning}: The tool suggests initial research strategies in accordance with the ACM SIGSOFT Empirical Standards\cite{ralph2020empirical}, which are refined based on the researcher's focus, to provide a detailed methodological approach.
    \item \textbf{Adaptability and Contextual Relevance}: Acknowledging the diverse nature of questions in disruptive technology research, MyResearchPlaybook adapts its recommendations to various contexts, including empirical studies, literature reviews, and case studies.
\end{itemize}

In summary, MyResearchPlaybook stands as an innovative addition to our research methodology, introducing a structured, rigorous approach to investigating disruptive technologies in software engineering. Its deployment aims to enhance the analytical depth and breadth of our studies, ensuring that our contributions are not only technologically advanced but also socio-technically informed and ethically sound.

\subsection{Limitations of MyResearchPlaybook}

While MyResearchPlaybook represents a significant advancement in AI-assisted research in the field of software engineering and disruptive technologies, it is imperative to acknowledge its limitations. This cautionary note serves to highlight areas where reliance on MyResearchPlaybook may require supplementation with human expertise and critical analysis.

\begin{itemize}
    \item \textbf{Dependence on Pre-Existing Knowledge}: MyResearchPlaybook operates within the bounds of its training data. This limitation means that the tool may not be aware of the very latest research developments or emerging technologies that have arisen post-training.
    
    \item \textbf{Lack of Domain-Specific Depth}: While MyResearchPlaybook is tailored for research in disruptive technologies, its understanding may not match the depth of knowledge of a seasoned expert in a specific sub-field. It is a generalist tool and may not fully grasp highly specialized or niche areas.
    
    \item \textbf{Potential for Bias}: Like any AI tool, MyResearchPlaybook may inadvertently reflect biases present in its training data. Researchers should be cautious of these potential biases and critically evaluate the suggestions and insights offered by the tool.
    
    \item \textbf{Absence of Creative Insight}: AI models, including MyResearchPlaybook, excel at analyzing and synthesizing existing information but do not possess the creative or innovative capabilities of human researchers. They cannot generate novel theories or hypotheses that have not been previously conceptualized.
    
    \item \textbf{Need for Human Interpretation}: The tool's outputs require interpretation and contextualization by knowledgeable researchers. MyResearchPlaybook can assist in identifying patterns and generating hypotheses, but the ultimate analysis and decision-making should rest with human researchers.
    
    \item \textbf{Ethical Considerations}: Ethical considerations, particularly in the field of disruptive technologies, often require nuanced understanding and judgement that may be beyond the scope of an AI model like MyResearchPlaybook.
\end{itemize}

Researchers are advised to use MyResearchPlaybook as a complement to, rather than a replacement for, their expertise and critical thinking skills. The tool should be viewed as one component in a broader methodological toolkit, offering support but not supplanting the essential human elements of curiosity, creativity, and ethical judgement in research. 


\end{document}